\documentclass[review]{elsarticle}

\usepackage{lineno,hyperref}
\usepackage{amsmath,amsfonts,amsthm,amssymb,graphicx,epstopdf,dcolumn,bm}
\usepackage{float,placeins,adjustbox}

\newtheorem{theorem}{Theorem}[section]

\newtheorem{definition}{Definition}[section]

\modulolinenumbers[5]

\journal{Journal of \LaTeX\ Templates}

\begin{document}

\begin{frontmatter}

\title{Dynamical system of cosmological models for different possibilities of $G$ and $\rho_{\Lambda}$}
\tnotetext[mytitlenote]{Fully documented templates are available in the elsarticle package on \href{http://www.ctan.org/tex-archive/macros/latex/contrib/elsarticle}{CTAN}.}


\author[Sonia]{Chingtham Sonia\corref{mycorrespondingauthor}}
\cortext[mycorrespondingauthor]{Corresponding author}
\ead{chingtham.sonia19@gmail.com}
\address[Sonia]{Department of Mathematics, National Institute of Technology Manipur, Imphal, 795004, India}

\author[Suren]{S. Surendra Singh}
\ead{ssuren.mu@gmail.com}
\address[Suren]{Department of Mathematics, National Institute of Technology Manipur, Imphal, 795004, India}

\begin{abstract}
The present paper deals with the dynamics of spatially flat Friedmann-Lema\^{i}tre-Robertson-Walker $(FLRW)$ cosmological model with a time varying cosmological constant $\Lambda$ where $\Lambda$ evolves with the cosmic time $t$ through the Hubble parameter $H$, that is, $\Lambda(H)$. We consider that the model dynamics has a reflection symmetry $H \rightarrow -H $ with $\Lambda(H)$ expressed in the form of Taylor series with respect to $H$. Dynamical systems for three different cases based on the possibilities of gravitational constant $G$ and the vacuum energy density $\rho_{\Lambda}$ have been analysed. In case I, both $G$ and $\rho_{\Lambda}$ are taken to be constant. We analyse stability of the system by using the notion of spectral radius, behavior of perturbation along each of the axis with respect to cosmic time and Poincar\'{e} sphere. In case II, we have dynamical system analysis for $G$=constant and $\rho_{\Lambda} \neq $constant where we study stability by using the concept of spectral radius and perturbation function. In case III, we take $G \neq$ constant and $\rho_{\Lambda} \neq$ constant where we introduce a new set of variables to set up the corresponding dynamical system. We find out the fixed points of the system and analyse the stability from different directions: by analysing behaviour of the perturbation along each of the axis, Center Manifold Theory and stability at infinity using Poincar\'{e} sphere respectively. Phase plots and perturbation plots have been presented. We deeply study the cosmological scenario with respect to the fixed points obtained and analyze the late time behavior of the Universe. Our model agrees with the fact that the Universe is in the epoch of accelerated expansion. The EOS parameter $\omega_{eff}$, total energy density $\Omega_{tt}$ are also evaluated at the fixed points for each of the three cases and these values are in agreement with the observational values in \cite{Planck}.
\end{abstract}


\end{frontmatter}

\section{Introduction}

In the past two decades many researchers have put tremendous efforts to develop and improve the plethora of theoretical models that explain the accelerated expansion of our Universe. Astrophysical measurements that reveal such a phenomenon put into the quest to give convincing theoretical explanations from various possible directions \cite{Riess1998, Perlmutter, Riess1999, Spergel, Tegmark, Abazajian2004, Abazajian2005, Spergel2007, Komatsu2009, Komatsu2011, Hinshaw}. The dark energy model is one such proposed model that attributes the expansion phenomenon to an energy component with negative pressure so called dark energy which dominates the universe at late time. The simplest type of dark energy is the cosmological constant \cite{Peebles}. In this context of accelerated expansion the theory of general relativity $(GR)$ modified by a cosmological constant term $\Lambda$, which is known as the famous $\Lambda$ CDM model is one of the most popular one \cite{Ostriker}. But, despite its fine agreement with the observation data, there are two major issues that have driven our young minds to focus sharply on some modifications to the assumed $\Lambda$ CDM model, namely, "the cosmological constant problem" which deals with the discrepancy between theoretical and expected values of the cosmological constant \cite{Weinberg, Nobbenhuis, Padmanabhan}; and "the cosmic coincidence problem \cite{Hamid Shabani}. To mend up these issues, running $\Lambda$ cosmological models have been developed.\\

Shapiro et al.\cite{Shapiro2002, Shapiro2003, Shapiro2005, Shapiro2009} made the first development regarding the scaling evolution of the cosmological constant. Among the running cosmological constant models that have been proposed, it is worthy enough to mention the time dependent cosmological constant motivated by quantum field theory \cite{Shapiro2009, Bonanno, Urbanowski}, $\Lambda(t)$ cosmology induced by a slowly varying Elko field \cite{Pereira}, a running vacuum in the context of supergravity \cite{Mavromatos}, etc. In Newtonian gravity, without any requirement of further constraints to be satisfied we can explicitly write the time variation of $G$. But in GR there are other constraints to be satisfied. For instance if we assume that the ordinary energy-momentum conservation law holds then there should not be any variation in the gravitational coupling with respect to the space time or otherwise the ordinary energy-momentum conservation law will be violated \cite{Hooman Moradpour, Canuto}. In the light of Dirac's idea \cite{Dirac1937, Dirac1938, Dirac1937p} which propose that some of the fundamental constants cannot remain constant forever, it is essential to do some modifications in GR field equations \cite{Chadrasekhar;Kothari, Saibal Ray} if we are to consider this running cosmological constant term. In this regard, studying the cosmic scenario with varying $G$ needs modified field equations as well as modified conservation laws. We can mention Brans-Dicke theory where there are modifications of GR with a varying $G$ without violating the ordinary energy-momentum conservation law \cite{Barrow, Brans, R. H. Dicke}. There are many other models that employ varying $G$ theories that give a better understanding of the Universe regarding its late time behavior and nature \cite{R. H. Dicke, Nojiri, Allemandi, Koivisto, Bertolami, Harko, Carloni, Boehmer, Rastall, Moradpour, De, Josset T, Das, Lin}. As there are no rigorous proves that indicate whether the cosmological constant is running or not \cite{Lin}, one can study the cosmological implications of different possible theoretical assumptions of $\Lambda$ term. Motivated by the quantum field theory \cite{Shapiro2003, Shapiro2005, Espana-Bonet} and some theoretical motivations \cite{Shapiro2009, Bonanno} about the varying $\Lambda$ form. Aleksander Stachowski, Marek Szydtowski \cite{Aleksander} have also studied the dynamics of cosmological models with various forms of $\Lambda(t)$.\\

In this paper, we consider a running vacuum model which evolves in power series of $H$. Our aim is to set up dynamical systems out of the cosmological field equations by introducing new set of variables and study the stability of the systems in the light of cosmological implications of the system. Based on the possibilities of the gravitational constant $G$ and the vacuum energy density $\rho_{\Lambda}$, we develop different dynamical system for three cases and analyze the stability through different approaches by finding respective fixed points. The cosmological scenario associated with each fixed point has been discussed in detail. We arrange the paper in the following ways. In section 1 we have given the introduction part, in section 2, we give preliminaries that provides a brief introduction on dynamical systems approach to cosmology with some definitions and theorems which will be required to understand the subsequent analysis in the paper. In section 3, we have three cases. In case I of section 3 we show the setting up of cosmological equations and dynamical system analysis where both $G$ and $\rho_{\Lambda}$ are taken to be constant which is the case of standard $\Lambda$ CDM cosmology. Under Case I we have three subsections based on analysis using spectral radius, perturbation function and stability at infinity using Poincar\'{e} sphere.  We present, in Case II, the model dynamics where $G$=constant and $\rho_{\Lambda} \neq $constant. Under case II, we have two subsections based on analysis through spectral radius and using perturbation along each of the axis with respect to increase in cosmic time. In Case III we have dynamical system analysis where $G \neq$ constant and $\rho_{\Lambda} \neq$ constant. Under Case III we present three subsections on the basis of analysing stability by the use of perturbation function, Center Manifold Theory and Poincar\'{e} sphere. In section 4 we give conclusion of our study. Stability analysis for each of the cases at the respective fixed points is presented and their corresponding cosmological implications along with the evaluation of various cosmological parameters at the respective fixed points are also obtained.

\section{Preliminaries}

Dynamical system is a mathematical system that describes the time dependence of the position of a point in the space that surrounds it, termed as ambient space. Here, we are approaching towards the system through an autonomous system of ordinary differential equations, (ASODE). ASODE is a system of ordinary differential equations which does not depend explicitly on time. S. Surendra et al. \cite{suren} have also used this approach to study cosmological models in the presence of a scalar field using different forms of potential. From \cite{suren} we can also notice that in three dimensional dynamical system we can analyse stability by analysing the nature of perturbation along each of the axis. A dynamical system is generally written in the form of the following \cite{Wiggins}:\\

 \begin{equation}\label{fd}
  \dot{x}=f(x),
 \end{equation}
 where $x=(x_{1},x_{2},......x_{n})$ is an element of state space $X\subseteq \mathbb{R}^{n}$ and the function $f:X \rightarrow X$.

The overhead dot denotes the derivative with respect to cosmic time, $t$. The function $f(x)$ is such that $f(x)=(f_{1}(x), f_{2}(x),...f_{n}(x))$ which can be viewed as a vector field in $\mathbb{R}^{n}$.

\begin{definition}\cite{Sebastian}~\\
\textbf{Fixed Point}: The point $x=x_{o}$ of the state space $X \subset \mathbb{R}^{n}$ is said to be a fixed point of the autonomous equation $\dot{x}=f(x)$ if and only if $f(x_{o})=0$.
\end{definition}

\begin{definition}\cite{Sebastian}~\\
  \textbf{Stable Fixed Point}: A fixed point $x_{o} $ of a dynamical system represented by $\dot{x}=f(x)$ is called a stable fixed point if for every $\epsilon >0$ there exist $\delta$ such that if $\psi(t)$ is any solution of the system satisfying $\parallel \psi(t_{o})-x_{o} \parallel < \delta$, then the solution $\psi(t)$ exists for all $t\geq t_{o}$ and it satisfies $ \parallel \psi(t)-x_{o} \parallel < \epsilon$ for all $t\geq t_{o}$.
\end{definition}

\begin{definition}\cite{Rafael}~\\
\textbf{Local Stability}: Let $g:I \rightarrow I$ be a map and $x_{o}$ be a fixed point of $g$, where $I$ is an interval of real numbers. Then
\begin{description}
  \item[(i)] the fixed point $x_{o}$ is said to be locally stable if, for any $\epsilon>0$, there exists $\delta>0$ such that, for all $x\in I$ with $\mid x-x_{o}\mid <\delta$, we have $\mid g^{n}(x)-x_{o}\mid <\epsilon$, for all $n \in \mathbb{N}$. Otherwise, the fixed point $x_{o}$ will be called unstable;\\

  \item[(ii)] the fixed point $x_{o}$ is said to be attracting if there exists $\zeta>0$ such that $\mid x-x_{o}\mid< \zeta$ implies $lim_{n \rightarrow \infty}g^{n}(x)=x_{o}$;\\

  \item[(iii)] the fixed point $x_{o}$ is said to be locally asymptotically stable if it is both stable and attracting. If in the previous item $\zeta=\infty$, then $x_{o}$ is said to be globally asymptotically stable.
\end{description}
\end{definition}

\begin{definition}~\\
  \textbf{Hyperbolic point}: A fixed point $x=x_{o} \in X \subset \mathbb{R}^{n}$ of the system $\dot{x}=f(x)$ is said to be a hyperbolic fixed point if none of the eigenvalues of the Jacobian matrix at $x_{o}$ ,$J(x_{o})$ have zero real part, otherwise the point is called non-hyperbolic.
\end{definition}

\begin{definition}\label{JMdef}~\\
\textbf{Jacobian matrix of dynamical system at a fixed point}: The Jacobian matrix of the dynamical system given in \eqref{fd} at a fixed point $x_{o}$ is given by \\
\[ J_{x_{o}} = \begin{bmatrix}
            \frac{\delta f_{1}}{\delta x_{1}}& \frac{\delta f_{1}}{\delta x_{2}}& . & . & . &\frac{\delta f_{1}}{\delta x_{n}}  \\
          . &~~~~~~~.&&&&.\\
          .&&.&&& .\\
          . && &.& & .\\
          \frac{\delta f_{n}}{\delta x_{1}}    &\frac{\delta f_{n}}{\delta x_{2}}&. & .& .& \frac{\delta f_{n}}{\delta x_{n}}
           \end{bmatrix} \]
where $\frac{\delta f_{i}}{\delta x_{i}}$, $i=1,2,...,n$ denotes the first partial derivative of $f_{i}$ with respect to the $i^{th}$ component $x_{i}$ of the element $x=(x_{1},x_{2},...x_{n})\in X \subseteq \mathbb{R}^{n}$ .
\end{definition}

Linear stability theory is one of the simplest method used to understand the dynamics of a system near a fixed point. In Linear stability theory the function $f$ is assumed to be sufficiently regular so that we can linearise the system around its fixed point. The eigenvalues of the Jacobian matrix at a fixed point play an important role in studying the stability of the fixed point.

For hyperbolic fixed points if all the eigenvalues of $J_{x_{o}}$ have positive real parts, then $x_{o}$ acts as a \emph{repeller} and it is unstable as all the trajectories closed enough to it are repelled from it. $x_{o}$ is stable when all the eigenvalues of $J_{x_{o}}$ have negative real parts.  Here $x_{o}$ is called as \emph{attractor} and it attracts all nearby trajectories towards it. If at least two eigenvalues have real parts with opposite sign then, $x_{o}$ behaves as a saddle fixed point which attracts trajectories in some directions and repels along other directions. \\

If at least one of the eigenvalues of the Jacobian matrix at a fixed point $x_{o}$ have zero real part then we can not do stability analysis by using eigenvalues of the Jacobian matrix. Such a fixed point is referred to as non-hyperbolic fixed point. To analyse stability of such fixed points we need a better approach other than the linear stability analysis like Center manifold theory, perturbation function, Lyapunov stability. Centre manifold theory is the most popular method which reduces the dimensionality of the system and determines the stability of the critical points of the parent system according as the stability of the reduced system. Wiggins \cite{Wiggins} and Carr \cite{Carr} have discussed the centre manifold theory in detail.

The eigenvalues of the Jacobian matrix $J$ with order $n\times n$ given in Definition \ref{JMdef} will have $n$ eigenvalues.
The eigenvectors of $J$ associated to the eigenvalues with negative real part spans a vector space called stable space, $J^{s}$ and the eigenvectors associated with positive real part spans a vector space called the unstable space, $J^{u}$. Similarly $J^{c}$ represents the vector space spanned by the eigenvectors associated with zero real part. Here, the superscript $s,u,c$ denote the dimensions of the respective vector spaces. Also the spaces $J^{s},J^{u}$ and $J^{c}$ are the subspaces of $\mathbb{R}^{n}$. The space $\mathbb{R}^{n}$ can be written as the direct sum of these three subspaces, that is, $\mathbb{R}^{n}=J^{s}\oplus J^{u}\oplus J^{c}$. These results have been detailed in Carr's book \cite{Carr}, Elaydi's book \cite{Elaydi} and Zhang's book \cite{Zhang}. If at least one eigenvalue of $J$ at a fixed point $x_{o}$ has positive real part then $x_{o}$ will be unstable whether it is hyperbolic or not. But if $x_{o}$ is non-hyperbolic and no eigenvalues has positive real part, then we can use Center manifold theory to determine stability of the fixed point.

Let us consider a two dimensional dynamical system. Using a suitable coordinate transformation we can rewrite any system of the form \eqref{fd} as follows:

\begin{equation}\label{AB}
\left.\begin{array}{c}
    \dot{\textbf{x}}=A\textbf{x} + f(\textbf{x,y}), \\
    \dot{\textbf{y}}=B\textbf{y} + g(\textbf{x,y}),
   \end{array}\right\}
\end{equation}
where $A$ is a $c\times c$ matrix having eigenvalues with zero real parts, $B$ is an $s \times s$ matrix having eigenvalues with negative real parts and $(\textbf{x,y})\in J^{c} \times J^{s}$. The functions $f$ and $g$ satisfy the following:

\begin{eqnarray}\label{fg}
  f(0,0) &=& 0 \\
  g(0,0) &=& 0 \\
  \nabla f(0,0)&=& 0 \\
  \nabla g(0,0) &=& 0.
\end{eqnarray}

\begin{definition}\cite{Carr}~\\
\textbf{Centre Manifold}: A geometrical space $M^{c}(0)$ is a centre manifold for \eqref{AB} if it can be locally represented as
\begin{equation}\label{cm}
  M^{c}(0)=\{(\textbf{x,y})\in J^{c}\times J^{s} | \textbf{y}=h(\textbf{x}),|\textbf{x}|<\delta, h(0)=0,\nabla h(0)=0\},
\end{equation}
for a sufficiently regular function $h(\textbf{x})$ on $J^{s}$ and $\delta$ however small it may be. The proofs of the existence of the centre manifold for the system \eqref{AB} is also provided in \cite{Carr}  and he has given the dynamics of the system \eqref{AB} restricted to the centre manifold as follows:

\begin{equation}\label{cm ext}
  \dot{\textbf{v}}=A\textbf{v}+f(\textbf{v},h(\textbf{v})),
\end{equation}
for sufficiently small $\textbf{v} \in \mathbb{R}^{c}$.
\end{definition}

\begin{theorem}\label{Theorem1}\cite{Poincare}
Consider a flow defined by a dynamical system on $\mathbb{R}^{2}$
\begin{equation}\label{Polynomial1}
  \left.\begin{array}{c}
  \dot{x} = P_{1}(x,y), \\
  \dot{y} = P_{2}(x,y),
\end{array}\right\}
\end{equation}
where $P_{1}$ and $P_{2}$ are polynomial functions of $x$ and $y$. Let $P_{1m}$ and $P_{2m}$ denote the $m^{th}$ degree term in $P_{1}$ and $P_{2}$ respectively. Then, the critical points at infinity for the $m^{th}$ degree polynomial system \eqref{Polynomial1} occur at the points $(X,Y,0)$ on the equator of the Poincar\'{e} sphere where

\begin{center}
$X^{2}+Y^{2}=1$ and $XP_{2m}(X,Y)-YP_{1m}(X,Y)=0$,
\end{center}
or equivalently at the polar angle $\theta_{j}$ and $\theta_{j}+\pi$ satisfying
\begin{center}
$G_{m+1}(\theta)\equiv cos\theta Q_{m}(cos\theta, sin\theta)-sin\theta P_{m}(cos\theta, sin\theta)=0$
\end{center}
This equation has at most $m+1$ pairs of roots $\theta_{j}$ and $\theta_{j}+\pi$ unless $G_{m+1}(\theta)$
is identically zero. If  $G_{m+1}(\theta)$ is not identically zero, then the flow on the equator of the Poincar\'{e} sphere is counter-clockwise at points corresponding
to polar angles $\theta$ where $G_{m+1}(\theta)>0$ and it is clockwise at points corresponding to polar angles $\theta$ where $G_{m+1}(\theta)<0$.
\end{theorem}

\begin{theorem}\label{Theorem2}\cite{Poincare}
The flow defined by \eqref{Polynomial1} in a neighborhood of any
critical point of \eqref{Polynomial1} on the equator of $S^{2}$, except the points $(0, \pm 1, 0)$, is
topologically equivalent to the flow defined by the following system
\begin{eqnarray*}
   && \pm \dot{y}=yz^{m}P_{1}(\frac{1}{z}, \frac{y}{z})-z^{m}P_{2}(\frac{1}{z}, \frac{y}{z}), \\
   && \pm\dot{z}=z^{m+1}P_{1}(\frac{1}{z}, \frac{y}{z}),
\end{eqnarray*}
the signs being determined by the flow on the equator of $S^{2}$ as determined in Theorem \ref{Theorem1}.
\end{theorem}

\begin{theorem}\label{Theorem3}\cite{Poincare}

Let us consider a flow in $\mathbb{R}^{3}$ defined by
\begin{equation}\label{Polynomial}
  \left.\begin{array}{c}
  \dot{x} = P_{1}(x,y,z), \\
  \dot{y} = P_{2}(x,y,z), \\
  \dot{y} = P_{3}(x,y,z),
\end{array}\right\}
\end{equation}
where $P_{1}$, $P_{2}$ and $P_{3}$ are polynomial functions of $x$, $y$, $z$ of maximum degree $m$.\\

The critical points at infinity for the $m^{th}$ degree polynomial
system \eqref{Polynomial} occur at the points $(X, Y, Z, 0)$ on the equator of the Poincar\'{e}
sphere $S^{3}$ where $X^{2}+Y^{2}+Z^{2}=1$ and
\begin{eqnarray*}
  XP_{2m}(X,Y,Z)-YP_{1m}(X,Y,Z) &=& 0, \\
  XP_{3m}(X,Y,Z)-ZP_{1m}(X,Y,Z) &=& 0, \\
  YP_{3m}(X,Y,Z)-ZP_{2m}(X,Y,Z) &=& 0,
\end{eqnarray*}
where $P_{1m}$, $P_{2m}$ and $P_{3m}$ denote the $m^{th}$ degree terms in $P_{1}$, $P_{2}$ and $P_{3}$ respectively.
\end{theorem}

\begin{theorem}\label{Theorem4}\cite{Poincare}
The flow defined by the system \eqref{Polynomial} in a neighborhood of $(\pm 1, 0, 0, 0) \in S^{3}$ is topologically equivalent to the flow defined by the system:
\begin{eqnarray*}
&& \pm\dot{y}=yw^{m}P_{1}(\frac{1}{w},\frac{y}{w},\frac{z}{w})-w^{m}P_{2}(\frac{1}{w},\frac{y}{w},\frac{z}{w}), \\
&& \pm\dot{z}=zw^{m}P_{1}(\frac{1}{w},\frac{y}{w},\frac{z}{w})-w^{m}P_{3}(\frac{1}{w},\frac{y}{w},\frac{z}{w}), \\
&& \pm\dot{w}=w^{m+1}P_{1}(\frac{1}{w},\frac{y}{w},\frac{z}{w}).
\end{eqnarray*}
\end{theorem}

\section{Dynamical system analysis for different possibilities of $G$ and $\rho_{\Lambda}$}

In this section we present the dynamical system analysis when $G=$constant and $\rho_{\Lambda}$=constant. This is a standard model and we present it as case I of our analysis.

 {\bf Case I: Dynamical system analysis when $G=$constant and $\rho_{\Lambda}$=constant }\renewcommand {\theequation}{\arabic{equation}}\\

The Einstein field equations in the presence of cosmological constant $\Lambda$ are given by

\begin{equation}\label{EFE}
\left.\begin{array}{c}
  R_{\mu\nu}-\frac{1}{2}g_{\mu\nu}R = 8\pi G( T_{\mu\nu}+g_{\mu\nu}\rho_{\Lambda}), \\
  R_{\mu\nu}-\frac{1}{2}g_{\mu\nu}R = 8\pi G \textbf{T}_{\mu\nu},
\end{array}\right\}
\end{equation}
where $T_{\mu\nu}$ is the ordinary energy-momentum tensor, $ \textbf{T}_{\mu\nu}\equiv T_{\mu\nu}+g_{\mu\nu}\rho_{\Lambda}$ is the modified energy-momentum tensor and $\rho_{\Lambda}=\frac{\Lambda}{8\pi G}$ is the vacuum energy density in the presence of $\Lambda$. \\

We assume that the universe is filled with a perfect fluid with velocity four-vector field $V_{\mu}$. With this consideration we have $T_{\mu\nu}=-p_{m}g_{\mu\nu}+(\rho_{m}+p_{m})U_{\mu}U_{\nu}$, where $\rho_{m}$ is the density of matter-radiation and $p_{m}=(\gamma-1)\rho_{m}$ is the corresponding pressure. In the similar way, the modified energy-momentum tensor can be expressed as

\begin{equation}
  \textbf{T}=-p_{tt}g_{\mu\nu}+(\rho_{tt}+p_{tt})U_{\mu}U_{\nu},
\end{equation}

where $p_{tt}=p_{m}+p_{\Lambda}$, $\rho_{tt}=\rho_{m}+\rho_{\Lambda}$ and $p_{\Lambda}=-\rho_{\Lambda}$ is the associated pressure in the presence of $\Lambda$.
 With this substitution in the above expression we have

 \begin{equation}\label{MEMT}
  \textbf{T}=(\rho_{\Lambda}-p_{m})g_{\mu\nu}+(\rho_{m}+p_{m})U_{\mu}U_{\nu}
 \end{equation}

By assuming a spatially flat Friedmann-Lema\^{i}tre-Robertson-Walker(FLRW) metric along with the above modified energy-momentum tensor \cite{Carvalho, Lima-Maia, Spyros Basilakos, Joan}, we have the following gravitational field equations:
\begin{equation}\label{FE1}
 8\pi G \rho_{tt}\equiv8\pi G \rho_{m} + \Lambda=3H^{2},
\end{equation}

\begin{equation}\label{FE2}
 8\pi G p_{tt}\equiv8\pi G p_{m} - \Lambda=-2\dot{H}-3H^{2},
\end{equation}
where the overhead dot denotes the derivative with respect to the cosmic time $t$.

With the help of FLRW metric and the Bianchi identities by respecting the Cosmological Principle embodied in the FLRW metric we have the following generalized local conservation law:

 \begin{equation}\label{LCL}
  \dot{\rho_{m}}+\dot{\rho_{\Lambda}}+3H(\rho_{m}+p_{m}+\rho_{\Lambda}+p_{\Lambda})=0.
\end{equation}

 If we put $p_{\Lambda}=-\rho_{\Lambda}$ and $p_{m}=(\gamma-1)\rho_{m}$ in the above equation we have the following balanced conservation equation:

\begin{equation}\label{BCE}
    \dot {\rho_{m}} + 3\gamma H\rho_{m}=-\dot{\rho_{\Lambda}}.
 \end{equation}

 Since $\rho_{\Lambda}$ is taken to be constant the right hand side of the above equation vanishes to give the following equation:

 \begin{equation}\label{BCE0}
    \dot {\rho_{m}} + 3\gamma H\rho_{m}=0.
 \end{equation}

 Motivated by the work of Aleksander Stachowski et al.\cite{Aleksander}, let us consider that the cosmological constant $\Lambda$ evolves with time through the hubble parameter $H$ with $\Lambda(H)$ given in the form of Taylor series with respect to $H$.

 \begin{equation}\label{TS}
  \Lambda(H)=\sum_{n=1}^{\infty}\frac{1}{n!} \frac{d^{n}}{dH^{n}} \Lambda(H)|_{0}H^{n}
 \end{equation}

In addition let us consider that there is reflection symmetry with respect to $H$, that is , $H \rightarrow -H$. So, if the system has $\lambda(t)$ as its solution then, $\lambda(-t)$ is also a solution of the system. As a result only the terms containing even powers of $H$ are present in the above power series \eqref{TS}. Shapiro and Sol\`{a} \cite{Shapiro2009} have also considered in detail the contribution of only the even powers of Hubbble parameter to the time varying $\Lambda(t)$.

Using \eqref{TS} in \eqref{FE1}, we have

\begin{equation}\label{Hdot}
  2\dot H=\Lambda_{0}+(\alpha_{2}-3)H^{2}+\alpha_{4}H^{4}+...-8\pi G (\gamma-1) \rho_{m},
\end{equation}
where $\Lambda_{0}=\Lambda(H)|_{0}$ and $\alpha_{n}~'s, n=2i, i=1,2,...$ are the coefficients in the Taylor series expansion of $\Lambda(H)$ given by $\alpha_{n}=\frac{1}{n!}\frac{d^{n}\Lambda(H)}{dH^{n}}|_{0}, n=2i, i=1,2,...$

To set up the dynamical system we consider the following set of new variables: $x=(\frac{H}{8\pi G})^{2}$ and $y=\rho_{m}$. With this substitution we can expressed \eqref{Hdot} in terms of the new set of variables as follows:
\begin{equation}\label{Hdotx}
  2\dot H=8\pi G[C_{0}+\beta_{1}x+\beta_{2}x^{2}+...-(\gamma-1) y],
\end{equation}

where

$C_{o}=\frac{\Lambda_{o}}{8\pi G}$;
$\beta_{i}$= $\left\{
        \begin{array}{ll}
          (\alpha_{2}-3)8\pi G, & \hbox{$i=1$,} \\
          \alpha_{2j}(8\pi G)^{(2i-1)}, & \hbox{$j\geq 2$,}
        \end{array}
      \right.$
$i=1,2,...$

Using \eqref{Hdotx} and the newly introduced variables in the above field equations, we obtain the following set of ordinary differential equations which will represent the required dynamical system:

\begin{equation*}
  x'=\frac{dx}{d\Theta}= \frac{dx}{dt}\frac{dt}{d\Theta},
\end{equation*}
where $\Theta=\ln a$ denotes the logarithmic time with respect to the scale factor $a$. The overhead dash denotes the derivative with respect to $\Theta$ while the overhead dot denotes the derivative with respect to cosmic time $t$.
\begin{equation}\label{DS1x}
 x'=\frac{1}{8\pi G}(C_{0}+\beta_{1}x-(\gamma-1) y).
\end{equation}
 Here we consider only a few powers of $H$ beyond the term $C_{o}$ so as to ensure a better $\Lambda$CDM limit. All the other terms involving higher powers of $H$ are neglected as their contribution is completely negligible at present \cite{Harald}

\begin{equation*}
  y'=\frac{dy}{d\Theta}=\frac{dy}{dt}.\frac{dt}{d\Theta},
\end{equation*}
that is,
\begin{equation}\label{DS1y}
  y'=-3\gamma y.
\end{equation}

To analyse stability, firstly we need to find the fixed points of the system. For this we equate $x'=0$, $y'=0$, that is,
\begin{equation*}
  x'=\frac{1}{8\pi G}(C_{0}+\beta_{1}x-(\gamma-1)y)=0.
\end{equation*}
This implies
\begin{equation*}
  x=\frac{wy-C_{o}}{\beta_{1}},
\end{equation*}
where $\beta_{1}=(\alpha_{2}-3)8\pi G$ and $y'=-3\gamma y=0$.

This implies either $y=0$ or $\gamma=0$. We can also have $y\rightarrow 0$ in evaluating the fixed point. We need to observe both the possibilities and their implications to the evolving cosmological scenario. When $y=0$ in the expression of $x$ above we get $x=\frac{-C_{o}}{(\alpha_{2}-3)8\pi G}$. So the first fixed point we have obtained is $F_{1}=(\frac{-C_{o}}{(\alpha_{2}-3)8\pi G},0)$. Again when $\gamma=0$ then from \eqref{BCE0} we see that $\rho_{m}=$ constant. Let us suppose that $\rho_{m}=\xi$, that is, $y = \xi$. Then the second fixed point we have obtained for the case of $\gamma=0$ is $F_{2}=(\frac{-C_{o}-\xi}{(\alpha_{2}-3)8\pi G},y=\xi)$. When we consider $y\rightarrow 0$ we will obtain a special case of non-hyperbolic fixed points called a normally hyperbolic fixed point which is actually a set of non-isolated fixed points. For normally hyperbolic fixed points stability is decided by the sign of real part of the remaining eigenvalue even if one of the eigenvalue of the Jacobian matrix vanishes.
So when we choose $y \rightarrow 0$ then  we can write the fixed point as $F_{3}=(\frac{-C_{o}}{(\alpha_{2}-3)8\pi G},y\rightarrow 0)$. Now let us evaluate the Jacobian matrices $J_{F_{1}}$, $J_{F_{2}}$ and $J_{F_{3}}$ at the respective fixed points to study the stability of the system.

Let $f(x,y)=\frac{1}{8\pi G}(C_{0}+\beta_{1}x-(\gamma-1) y)$, $g(x,y)=-3\gamma y$.

The Jacobian matrix at the respective fixed points are given by\\

\begin{center}
$J_{F_{1}}=J_{F_{3}} =\left(
             \begin{array}{cc}
               f_{x} & f_{y} \\
               g_{x} & g_{y} \\
             \end{array}
           \right)
  =\left(
   \begin{array}{cc}
     \frac{\beta_{1}}{8\pi G} & \frac{-(\gamma-1)}{8\pi G} \\
     0 & -3\gamma \\
   \end{array}
 \right),
$
\end{center}

\begin{center}
$J_{F_{2}} = \left(
   \begin{array}{cc}
     \frac{\beta_{1}}{8\pi G} & \frac{1}{8\pi G} \\
     0 & 0 \\
   \end{array}
 \right)$,
\end{center}
where $\beta_{1}=(\alpha_{2}-3)8\pi G$.

The above matrices are upper triangular matrices. We all know that the eigenvalues of the Jacobian matrices are given by the diagonal entries. So, the eigenvalues of $J_{F_{1}}=J_{F_{3}}$ are $EV^{J_{1}}_{1}=EV^{J_{3}}_{1}=\frac{\beta_{1}}{8\pi G}=(\alpha_{2}-3)$, $EV^{J_{1}}_{2}=EV^{J_{3}}_{2}=-3\gamma$ and those of $J_{F_{2}}$ are $EV^{J_{2}}_{1}=EV^{J_{1}}_{1}=\frac{\beta_{1}}{8\pi G}=(\alpha_{2}-3)$, $EV^{J_{2}}_{2}=0$. The fixed points $F_{1}$ and $F_{3}$ are hyperbolic for $\gamma \neq 0$ as none of the eigenvalues vanishes. When $\gamma \neq 0$,  $EV^{J_{1}}_{1}$, $EV^{J_{3}}_{1}<0$ for $\alpha_{2}<3$ and $EV^{J_{3}}_{1}$,$EV^{J_{1}}_{2}<0$ for all $\gamma \in (0,2]$. As all the eigenvalues of $J_{F_{1}}$ and $J_{F_{3}}$ possess negative values for $\gamma \neq 0$, $\alpha_{2}<3$, $F_{1}$ and $F_{3}$  are stable fixed points. If $y \rightarrow 0$ is considered, though $EV^{J_{3}}_{2}=0$ $F_{3}$ is still stable as the remaining eigenvalue $(\alpha_{2}-3)$ is negative for $\alpha_{2}<3$. The fixed points $F_{1}$ and $F_{3}$ are stable and behaves as an attractor for $\alpha_{2}<3$. Fig. 1 and Fig. 2 shows the phase plot of $F_{1}$ for $\gamma=\frac{4}{3}$ and $\gamma=2$ respectively with $\alpha_{2}=2<3$ where all the nearby trajectories are attracted towards it. When $\alpha_{2}>3$, the eigenvalues of $J_{F_{1}}$ possess opposite signs which shows that $F_{1}$ behaves as a saddle fixed point. Fig. 3 shows the phase plot of the system for $\alpha_{2}=4>3$ where trajectories in some directions are attracted towards $F_{1}$ while trajectories along some other directions are repelled away from it. For the fixed point $F_{2}$ we see that $J_{F_{2}}$ is non-hyperbolic as one of the eigenvalues, namely, $EV^{J_{2}}_{2}=0$. For non-hyperbolic fixed point $F_{2}$ we can not analyse stability using the above linear stability theory. Since it is a two dimensional dynamical system we can use the notion of perturbation function and spectral radius of the Jacobian matrix for the non-hyperbolic fixed point $F_{2}$ to analyse the stability. In the subsequent paragraph we will show the stability analysis using these methods .\\

\textbf{A. } {\bf Stability analysis for $F_{2}$ using the concept of Spectral radius:}\renewcommand {\theequation}{\arabic{equation}}\\

 Let's rewrite the Jacobian matrix at the fixed point $F_{2}$ as follows:

\begin{center}
$J_{F_{2}}=\left(
   \begin{array}{cc}
     (\alpha_{2}-3) & \frac{1}{8\pi G} \\
     0 & 0 \\
   \end{array}
 \right).
$
\end{center}
Trace of $J_{F_{2}}$, $tr(J_{F_{2}})$= sum of eigenvalues= $ EV^{J_{2}}_{1} +EV^{J_{2}}_{2} $ =$(\alpha_{2}-3)$.\\
Determinant of $J_{F_{2}}$, $det(J_{F_{2}})$= product of eigenvalues= $EV^{J_{2}}_{1} \times EV^{J_{2}}_{2} =0$ .\\

The spectral radius of a matrix is the maximum of the absolute values of all the eigenvalues of the matrix. The stability of a fixed point $(x,y)$ of a dynamical system can be determined by the value of spectral radius of its Jacobian matrix evaluated at the fixed point. The notion of spectral radius in discussing stability of a fixed point has been given in detail in \cite{Elaydi}.

 The spectral radius of the above Jacobian matrix is given by
\begin{eqnarray}\label{SR}
  \sigma_{J_{F_{2}}} &=& max \{|\lambda| : \lambda ~is~ the~ eigenvalue \}, \nonumber\\
    &=&max\{ |\alpha_{2}-3|,0 \}, \nonumber\\
    &=&\bigg\{ \begin{array}{cc}
            \alpha_{2}-3, & \alpha_{2}>3, \\
           -(\alpha_{2}-3),  & \alpha_{2}<3.
          \end{array}
\end{eqnarray}

By theorem [\cite{Elaydi}, page 221], $F_{2}$ will be locally asymptotically stable if $\sigma_{J_{F_{2}}}<1$. And we can not determine stability when $\sigma_{J_{F_{2}}}=1$ and hence Centre manifold theory is the most viable way to analyse stability. With reference to [\cite{Elaydi} page 200], spectral radius will be less than unity if and only if
\begin{equation*}
  |tr(J_{F_{2}})|-1<det(J_{F_{2}})<1.
\end{equation*}

From the above arguments, $F_{2}$ is locally asymptotically stable for $3< \alpha_{2}<4$ or $2<\alpha_{2}<3$. It can be noted that we have assume $\alpha_{2}\neq3$ here so that we can study our system with fixed points in finite phase plane.\\

\textbf{B. } {\bf Stability analysis for $F_{2}$ using the concept of Perturbation function: }\renewcommand {\theequation}{\arabic{equation}}\\

To analyse stability in a simpler way we find perturbation function along each axis as a function of logarithmic time $\Theta$. It is noted that while studying perturbation along $x-$axis we assume $y=0$ as we are analysing only along $x-$axis. We can make the interval where $\alpha_{2}$ lies finer by analysing the stability from this side of perturbation function. Now to find the perturbation function we perturb the system by a small amount, that is, $x= \frac{-c_{o}-\xi}{(\alpha_{2}-3)8\pi G}+\eta_{x}$ and $y=\xi+ \eta_{y}$, where $\eta_{x}$ and $\eta_{y}$ represent small perturbations along $x$ and $y$ axes respectively. With these perturbed system, \eqref{DS1x} and \eqref{DS1y} takes the following form:

\begin{equation*}
  \eta_{x}' = \frac{1}{8\pi G}(C_{o}+\beta_{1}(\frac{-C_{o}-\xi}{(\alpha_{2}-3)8\pi G}+\eta_{x}).
\end{equation*}

Solving the above differential equation we obtain $\eta_{x}$ as a function of logarithmic time, $\Theta$ as follows:

 \begin{equation}\label{PDS1x}
  \eta_{x} = e^{(\alpha_{2}-3)\Theta} + \frac{\xi}{(\alpha_{2}-3)8\pi G}.
\end{equation}
Similarly,
\begin{equation}\label{PDS1y}
  \eta_{y} = \frac{C-\xi e^{3\gamma \Theta}}{e^{3\gamma \Theta}}.
\end{equation}

When $\alpha_{2}<3$, as $\Theta$ tends to infinity the perturbation along $x$-axis, $\eta_{x}$ evolves to a constant value which is $\frac{\xi}{(\alpha_{2}-3)8\pi G}$. In the above expression of $\eta_{y}$ if we consider $\Theta \rightarrow \infty $, we get $\frac{\infty}{\infty}$ form. So we can apply L Hospital's rule of finding limit in the expression of $\eta_{y}$ to obtain its limiting value as $-\xi$ for any value of $\gamma$ . We can also directly put $\gamma=0$ in \eqref{DS1y} to get $\eta_{y}'=0$ and obtain $\eta_{y}=$constant. But by doing so we won't be able to show the nature of $\eta_{y}$ in terms of $\Theta$ and further with \eqref{PDS1y} we can achieve the constant value towards which $\eta_{y}$ evolves in a finer way. As perturbation along both the axes evolve to a constant value when $\alpha_{2}<3$, we conclude that $F_{2}$ is stable for $\alpha_{2}<3$ and it is locally asymptotically stable for $2<\alpha<3$. If $\Phi=\{\alpha_{2}: \eta_{x}\rightarrow 0$ or a constant $\wedge$ $\eta_{y}\rightarrow 0$ or a constant $\wedge$ $ \eta_{z}\rightarrow $0 or a constant$\}$, then $F_{2}$ is stable for any $\alpha_{2} \in \Phi$ where $\Phi=(-\infty,3)$. The perturbation plots shown in Fig. 4 shows the variation of perturbation function along $y$ axis with respect to $\Theta$ for $F_{2}$. From Fig. 4 we see that when $\gamma=0$, $\eta_{y}$ becomes a constant function, but if $\gamma \neq 0$ then as $\Theta \rightarrow \infty$, $\eta_{y}$ takes $\frac{\infty}{\infty}$ form. So by applying L Hospital's rule as $\Theta \rightarrow \infty$, $\eta_{y}$ tends to $-\xi$ which is a constant value.  Fig. 5 shows that the perturbation along $x-$axis tends to a constant value, namely, $\frac{\xi}{(\alpha_{2}-3)8\pi G}$ when $\alpha_{2}<3$. In the plot shown in Fig. 5 we take $\xi=1$, $8\pi G=1$ and $\alpha_{2}=2.5<3$ to show that $\eta_{y}$ tends to $\frac{\xi}{(\alpha_{2}-3)8\pi G}=-2$ here. \\

In terms of the variables $x$ and $y$ we obtain the value of effective equation of state $\omega_{eff}$ and total energy density $\Omega_{tt}$ as follows:
\begin{eqnarray*}
  \omega_{eff} &=& \frac{p_{tt}}{\rho_{tt}}, \\
    &=& -1-\frac{\alpha_{2}-3}{3}-\frac{C_{o}}{24\pi Gx}+\frac{(\gamma-1)y}{24\pi Gx},
\end{eqnarray*}
where $\rho_{tt}=24\pi G x$ and $p_{tt}=(-24\pi Gx-\beta_{1}x-C_{o}+ (\gamma-1)y$;

\begin{equation*}
  \Omega_{tt}= \frac{\Lambda_{o}}{3(8\pi G)^{2}x}+\frac{\alpha_{2}}{3}+ \frac{y}{24\pi G x},
\end{equation*}
where vacuum energy density, $\Omega_{\Lambda}=\frac{\Lambda_{o}}{3(8\pi G)^{2}x}+\frac{\alpha_{2}}{3}$ and matter density, $\Omega_{m}=\frac{y}{24\pi G x}$.

At $F_{1}$ the value of effective equation of state parameter $\omega_{eff}$ is calculated as -1 which assures the presence of negative pressure in the existing cosmological scenario with the numerical value of $\Omega_{tt}$ as $\Omega_{tt} \approx 1$. Thus the presence of this late time attractor contributes to our model with an accelerated expansion phase of the Universe with $\omega_{eff}=-1$ and $\Omega_{tt}=0.99\approx 1$ which is in agreement with the observational data in \cite{Planck}. Also when we evaluated the above cosmological parameters at the fixed point $F_{2}$, for any value of $\alpha_{2}$ and $\xi$ we obtained $\omega_{eff}=-1$. The relative energy density at $F_{2}$ is found to be $\Omega_{tt}= 1$. The above results have been tabulated in TABLE I:

\begin{table}[H]
\caption{Table for case I ($G$=constant, $\rho_{\Lambda}$=constant)}
\label{table: Table1 }
\resizebox{\columnwidth}{!}{%
\begin{tabular}{|c|c|c|c|c|c|c|c|}
\hline
Fixed points & x & y & Type of fixed point & Eigen Values & $\omega_{eff}$ & $\Omega_{tt}$  & Behavior \\
\hline
$F_{1}$ & $\frac{-C_{o}}{\beta_{1}}$, & 0 & hyperbolic& $\alpha_{2}-3$, -3$\gamma$ & -1 &0.99 & stable for $\alpha_{2}<3$,$\gamma\neq 0$  \\
 & & & & & &$\approxeq$ 1 &,late time attractor \\
 & & & & & & & \\
& & & & & & &saddle point for $\alpha_{2}>3$, \\
 & where $\beta_{1}=(\alpha_{2}-3)8\pi G $ & & & & & &unstable. \\
 & & & & & & & \\
 & & & & & & & \\
$F_{2}$ & $\frac{(-C_{o}-\xi)}{\beta_{1}}$ & $\xi$ & non-hyperbolic &$(\alpha_{2}-3)$, 0 & & & stable for $\alpha_{2}<3$ ,\\
& & & & & & & \\
 & & & & & -1& 1 &locally asymptotically stable \\
& & & & & & & \\
 &where $\beta_{1}=(\alpha_{2}-3)8\pi G $ & & & & & &for $2<\alpha_{2}<3$ \\
 & & & & & & &\\
 & & & & & & & \\
$F_{3 }$ &$\frac{-C_{o}}{\beta_{1}}$  & $y \rightarrow 0$ & normally  & $(\alpha_{2}-3)$ ,0 &-1& $1 $ & stable for all $\gamma \in [0,2]$, \\
  & & &hyperbolic & & & & \\
 & where $\beta_{1}=(\alpha_{2}-3)8\pi G $ & & for $\gamma=0$& & & & behaves as late time attractor \\
 & & & & & & & for $\alpha_{2}<3$\\
\hline
\end{tabular}%
}
\end{table}

\begin{figure}[H]
\includegraphics[height=2in]{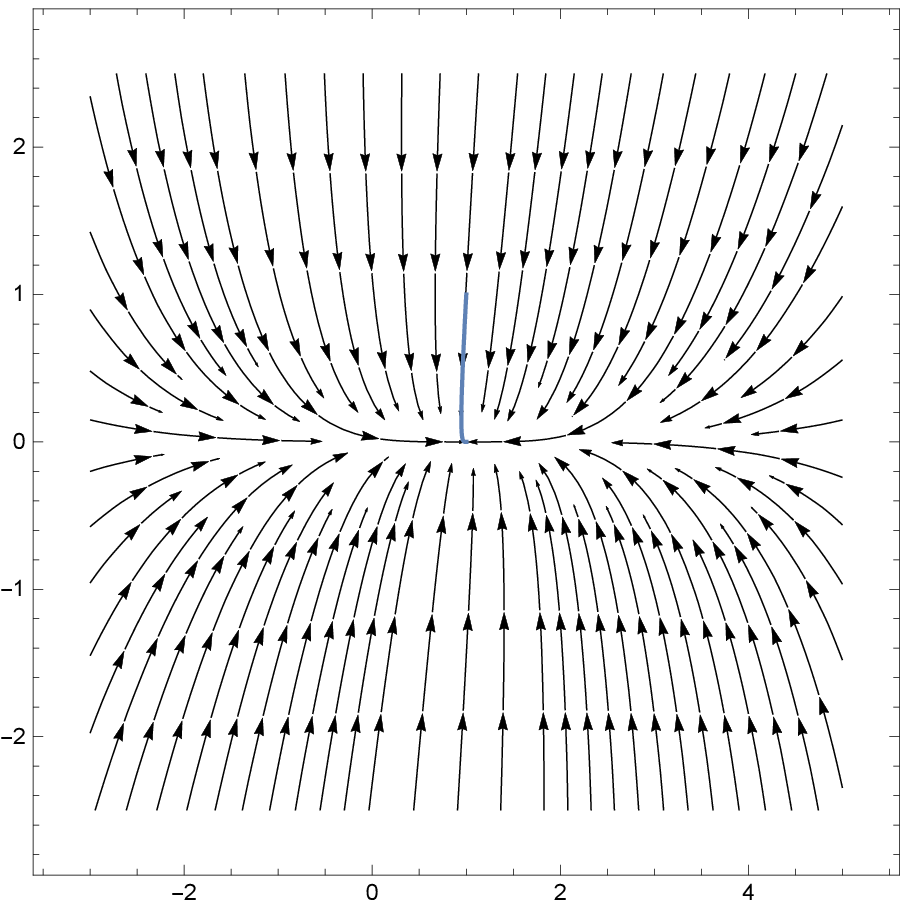}
\includegraphics[height=2in]{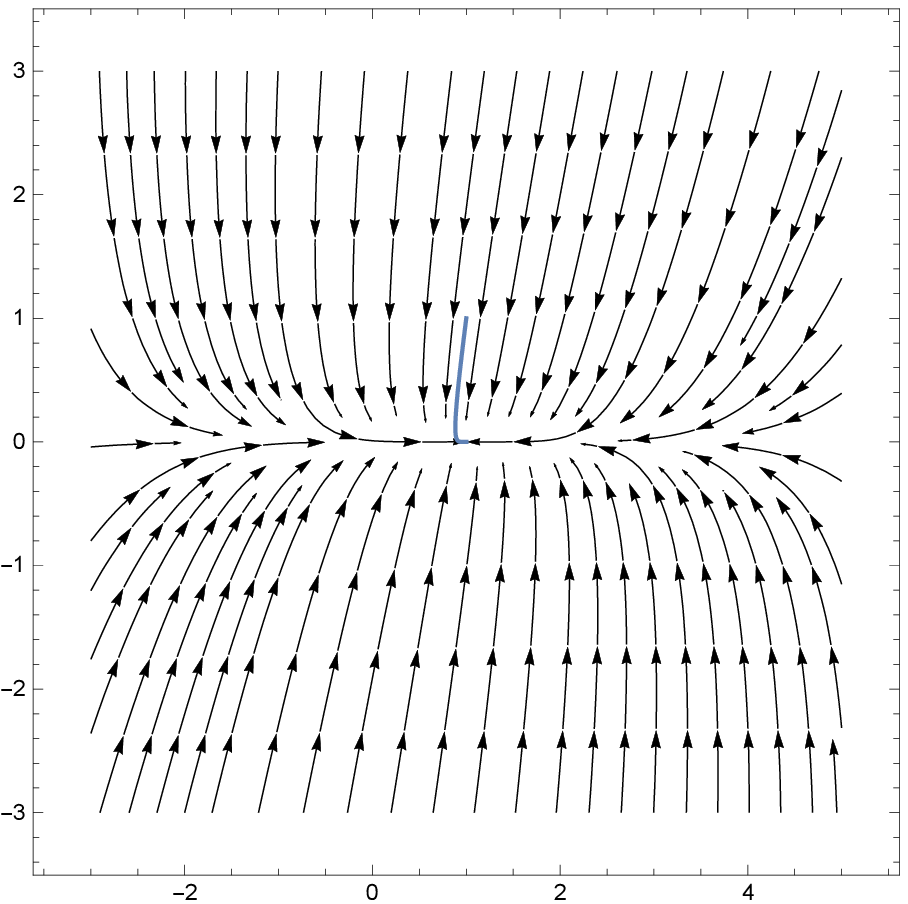}\\
\vspace{1mm}
~~~~~~~~~~~~~~~Fig. 1~~~~~~~~~~~~~~~~~~~~~~~~~~~~~~~~~~~~~~~~~~Fig. 2\\
\vspace{5mm}Fig. 1 shows the phase plot for $F_{1}$ at $\alpha_{2}=2<3$, $\gamma= \frac{4}{3}$, stable attractor.~~~Fig.2 shows the phase plot for stable $F_{1}$ at $\gamma=2$, $\alpha_{2}<3$.
\hspace{2cm} \vspace{6mm}
\end{figure}

\begin{figure}[H]
\includegraphics[height=2in]{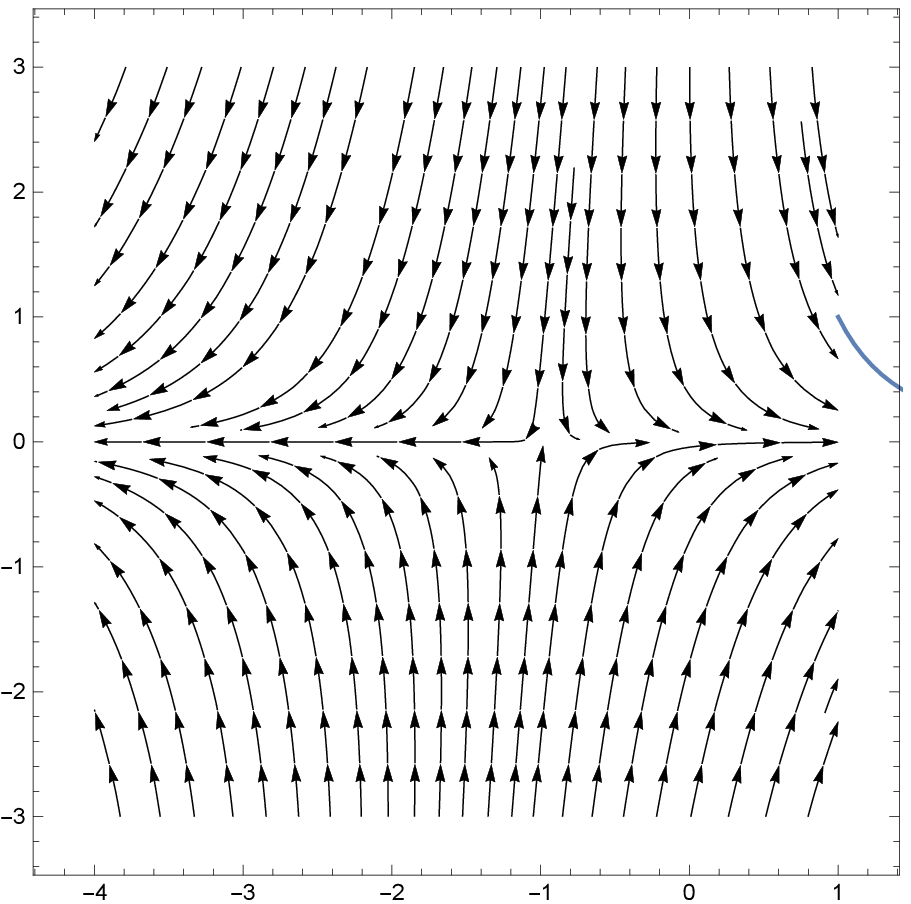}
\includegraphics[height=2in]{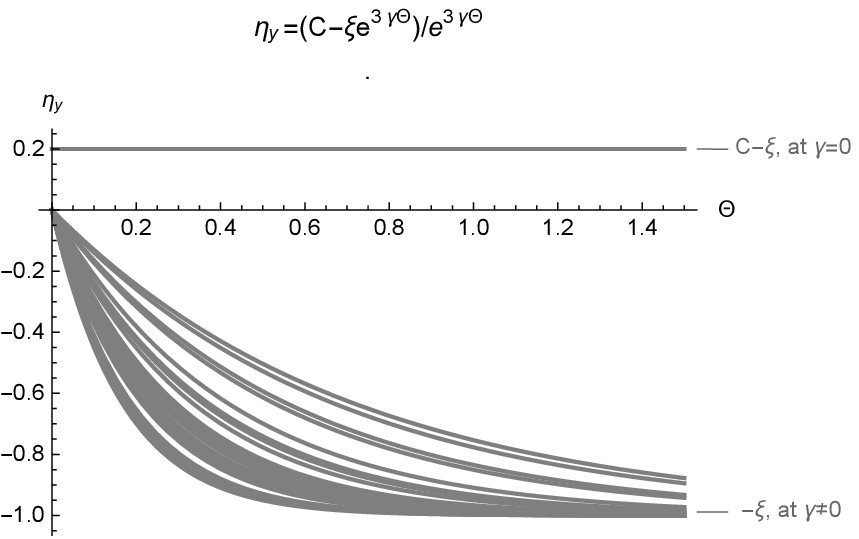}\\
\vspace{1mm}
~~~~~~~~~~~~~~Fig. 3~~~~~~~~~~~~~~~~~~~~~~~~~~~~~~~~~~~~~~~~~~Fig. 4\\
\vspace{5mm}Fig. 3 shows the phase plot for $F_{1}$ at $\alpha_{2}=4>3$ representing saddle point.~~~Fig.4 shows variation of $\eta_{y}$ with respect to $\Theta$ for $F_{2}$.
\hspace{2cm} \vspace{6mm}
\end{figure}

\begin{figure}[H]
\includegraphics[height=1.8in]{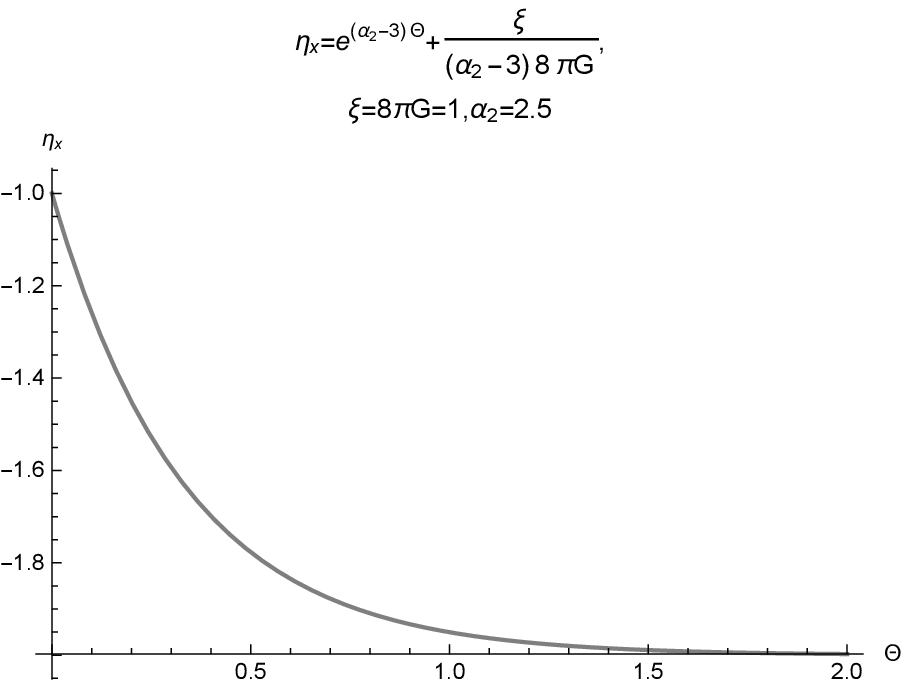}\\
\vspace{1mm}
~~~~~~~~~~~~~~~~~~~~Fig. 5\\
\vspace{5mm}Fig. 5 shows the variation of $\eta_{x}$ with respect to $\Theta$ for $F_{2}$.
\hspace{2cm} \vspace{6mm}
\end{figure}

\textbf{C. } {\bf  Stability at infinity and Poincar\'{e} sphere: }\renewcommand {\theequation}{\arabic{equation}}\\

The detail explanation of Poincar\'{e} sphere and behavior at infinity is given in \cite{Poincare}. By using stereographic projection we can study the behavior of trajectories far from origin by considering the so-called Poincar\'{e} sphere where we project from the center of the unit sphere $S^{2}=\{ (X,Y,Z)\in \mathbb{R}^{3}|X^{2}+Y^{2}+Z^{2}=1\}$ onto the $(x,y)-$plane tangent to $S^{2}$ at the north pole\cite{Poincare} by using the transformation of coordinates given by \\

\begin{equation}\label{xX}
  x=\frac{X}{Z}, ~ y=\frac{Y}{Z}.
\end{equation}

The equations defining $(X,Y,Z)$ in terms of $(x,y,z)$ are given by

\begin{equation*}
 X=\frac{x}{\sqrt{1+x^{2}+y^{2}}}, Y=\frac{y}{\sqrt{1+x^{2}+y^{2}}}, Z=\frac{1}{\sqrt{1+x^{2}+y^{2}}}.
\end{equation*}

The critical points at infinity are mapped on the equator of the Poincar\'{e} sphere. We consider the following flow in $\mathbb{R}^{2}$:

\begin{eqnarray}\label{2D1}
  x' &=& \frac{1}{8\pi G}(C_{o}+\beta_{1}x-(\gamma-1)y), \\
  y' &=& -3\gamma y.
\end{eqnarray}

Let $f(x,y)=\frac{1}{8\pi G}(C_{o}+\beta_{1}x-(\gamma-1)y)$, $g(x,y)=-3\gamma y$. The degree of this polynomial system is one and let $f_{1}$ and $g_{1}$ denotes the homogeneous polynomials in $f$ and $g$ of first degree, that is, $f_{1}=\frac{1}{8\pi G}(\beta_{1}x-(\gamma-1)y)$, $g_{1}=-3\gamma y$. In terms of the polar coordinates $r$, $\theta$ with $x=r\cos \theta$, $y=r\sin \theta$, we can express the above equations as

\begin{equation}\label{2D1polar1}
    r' = \frac{C_{o}cos\theta}{8\pi G}+r((3\gamma+\frac{\beta_{1}}{8\pi G})cos^{2}\theta-3\gamma+\frac{(\gamma-1)sin2\theta}{16\pi G}),
\end{equation}

\begin{equation}\label{2D1polar2}
    \theta'= \frac{-C_{o}sin\theta}{8\pi G}\frac{1}{r}-\frac{(3\gamma+\frac{\beta_{1}}{8\pi G})sin2\theta }{2}+\frac{(\gamma-1)sin^{2}\theta}{8\pi G}.
\end{equation}

Order of $r$ in \eqref{2D1polar1} as $r\rightarrow \infty$ is $\bar{i}=1$ and that of \eqref{2D1polar2} is $\bar{j}=0$. Let us denote $\bar{k}=\bar{i}-\bar{j}=1-0=1$. And using \eqref{2D1polar1}, we have
\begin{equation*}
  \lim_{r\rightarrow \infty}r'=\lim_{r\rightarrow \infty}\frac{dr}{d\Theta}=\infty \neq 0.
\end{equation*}

Then using Theorem \ref{Theorem1} we find $G_{2}(\theta)$ which is also equal to the highest power term in $r$ of the $\Theta'$ expression \cite{Roland}. \\

\begin{equation*}
  G_{2}(\theta)=\frac{-3\gamma sin2\theta}{2}-\frac{\beta_{1}sin2\theta}{16\pi G}+\frac{(\gamma-1)sin^{2}\theta}{8\pi G}.
\end{equation*}

Solving $\theta$ for which $G_{2}(\theta)=0$ we get $\theta=n\pi$, where $n=0,\pm1,\pm2,...$. So we can conclude that $G_{2}(\theta)$ is not identically equal to zero but it becomes zero in those directions where $\theta$ takes the value $n\pi$. Since $G_{2}(\theta)$ has at most 2 pairs of roots $\theta$ and $\theta+\pi$, the equator of the Poincar\'{e} sphere has finite number of fixed points located at $\theta$ such that $G_{2}(\theta)=0$, that is, at $\theta=0,\pi,\pi, 2\pi$ or equivalently $\theta=0, \pi$. At $\gamma=0, \frac{4}{3}$ and 2, $G_{2}(\theta)$ takes the following form: \\

\begin{equation}\label{G1}
  G_{2}(\theta)=\left\{
                  \begin{array}{ll}
                   \frac{-(\alpha_{2}-3)sin2\theta}{2}-\frac{sin^{2}\theta}{8\pi G}, & \hbox{$\gamma=0$;} \\
                   \frac{-(1+\alpha_{2})sin2\theta}{2}+\frac{sin^{2}\theta}{24\pi G}, & \hbox{$\gamma=\frac{4}{3}$;} \\
                    \frac{-(3+\alpha_{2})sin2\theta}{2}+\frac{sin^{2}\theta}{8\pi G}, & \hbox{$\gamma=2$.}
                  \end{array}
                \right.
\end{equation}

The flow on the equator of the Poincar\'{e} sphere is counterclockwise at points corresponding to polar angles $\{\theta: \theta<\tan^{-1}((3-\alpha_{2})8\pi G)\}$ where $G_{2}(\theta)>0$, for example $\theta=(2n\pi+\frac{\pi}{4}), n=0,\pm1,\pm2,...$ with $\alpha_{2}<3-\frac{1}{8\pi G}$. The flow is clockwise at points corresponding to polar angles $\{\theta: \theta>\tan^{-1}((3-\alpha_{2})8\pi G)\}$ where $G_{2}(\theta)<0$, for example $\theta=(2n+1)\frac{\pi}{2}$. For $\gamma=\frac{4}{3}$ the flow on the equator of the Poincar\'{e} sphere is counterclockwise at points corresponding to polar angles $\{\theta: \theta>\tan^{-1}((1+\alpha_{2})24\pi G)\}$ where $G_{2}(\theta)>0$ and the flow is clockwise at points corresponding to polar angles $\{\theta: \theta<\tan^{-1}((1+\alpha_{2})24\pi G)\}$ where $G_{2}(\theta)<0$. For $\gamma=2$ the flow is counterclockwise at points corresponding to polar angles $\{\theta: \theta>\tan^{-1}((3+\alpha_{2})8\pi G)\}$ where $G_{2}(\theta)>0$ and the flow is clockwise at those points corresponding to polar angles $\{\theta: \theta<\tan^{-1}((3+\alpha_{2})8\pi G)\}$ where $G_{2}(\theta)<0$.

By Theorem \ref{Theorem1}, the critical points at infinity for the system occur at the points $(X,Y,0)$ on the equator of the Poincar\'{e} sphere where $X^{2}+Y^{2}=1$ and
\begin{equation*}
  Xg_{1}(X,Y)-Yf_{1}(X,Y)=0,
\end{equation*}
where $f_{1}(x,y)=(\alpha_{2}-3)x-\frac{(\gamma-1)y}{8\pi G}$ and $g_{1}(x,y)=-3\gamma y$.
Using \eqref{xX}, the above equation becomes

\begin{equation}\label{clock}
  -3\gamma XY-(\alpha_{2}-3)XY+\frac{(\gamma-1)Y^{2}}{8\pi G}=0.
\end{equation}

Solving for $X$ and $Y$ from the above equations, we find that fixed point occurs at $(\pm1,0,0)$. Also we see from the expression in \eqref{clock} that for $\gamma=0$ the flow on the equator of $S^{2}$ is clockwise for $XY>0$ and counterclockwise for $XY<0$. For $gamma=\frac{4}{3}$, the flow on the equator of $S^{2}$ is clockwise for $XY>0$ and $-(1+\alpha_{2})XY>\frac{Y^{2}}{24\pi G}$ ; and the flow is counterclockwise for $XY<0$. For $gamma=2$, the flow on the equator of $S^{2}$ is clockwise for $XY>0$ and $-(3+\alpha_{2})XY>\frac{Y^{2}}{8\pi G}$ ; and the flow is counterclockwise for $XY<0$. Using Theorem \ref{Theorem2} The behavior in the neighbourhood of the critical point $(1,0,0)$ is topologically equivalent to the behavior of the following system,

\begin{equation}\label{a}
   y'=yzf(\frac{1}{z},\frac{y}{z})-zg(\frac{1}{z},\frac{y}{z}),
\end{equation}

\begin{equation}\label{b}
   z'=z^{2}f(\frac{1}{z},\frac{y}{z}).
\end{equation}

Putting the expressions of $f,g$ in \eqref{a} and \eqref{b} we get

\begin{equation}\label{c}
 y'=yz\frac{C_{o}}{8\pi G}+((\alpha_{2}-3)+3\gamma)y-\frac{(\gamma-1)y^{2}}{8\pi G},
\end{equation}

\begin{equation}\label{d}
 z'=\frac{C_{o}z^{2}}{8\pi G}+(\alpha_{2}-3)z-\frac{(\gamma-1)y}{8\pi G}.
\end{equation}

The Jacobian matrix of the above system is

\begin{center}
$J_{inf}(0,0)=\left(
  \begin{array}{cc}
    (\alpha_{2}-3)+3\gamma & 0 \\
    \frac{-(\gamma-1)}{8\pi G} & \alpha_{2}-3 \\
  \end{array}
\right)$
\end{center}

This is a lower triangular matrix. So the eigenvalues are given by the diagonal entries, that is,
$m_{1}=(\alpha_{2}-3)+3\gamma$ and $m_{2}=\alpha_{2}-3$. For $\gamma=0$, both $m_{1}$ and $m_{2}$ are negative for $\alpha_{2}<3$ and the critical point $(1,0,0)$ behaves as a stable attractor which represents the late time accelerated expansion phase of the Universe. For $\alpha_{2}>3$, both $m_{1}$ and $m_{2}$ are positive and the critical point $(1,0,0)$ behaves as an unstable repeller representing the inflationary epoch of the evolving Universe. Fig. 6 and Fig. 7 shows the phase plot of stable attractor as well as the unstable repeller respectively.

For $\gamma=\frac{4}{3}$, $m_{1}>0$ and $m_{2}<0$ when $\alpha_{2}<3$ and the critical point $(1,0,0)$ behaves as a saddle point which is unstable representing the matter dominated phase of the evolving Universe . When $\alpha_{2}>3$, both $m_{1}$ and $m_{2}$ are positive and the critical point $(1,0,0)$ behaves as an unstable repeller. For $\gamma=2$, the behavior is same as that of $\gamma=\frac{4}{3}$. Since the degree of $f(x,y)$ and $g(x,y)$ is odd, the behavior at the antipodal point $(-1,0,0)$ is exactly the same as the behavior at $(1,0,0)$. Fig. 8 and Fig. 9 show the phase plot for unstable saddle point and repeller respectively.

\begin{figure}[H]
\includegraphics[height=2in]{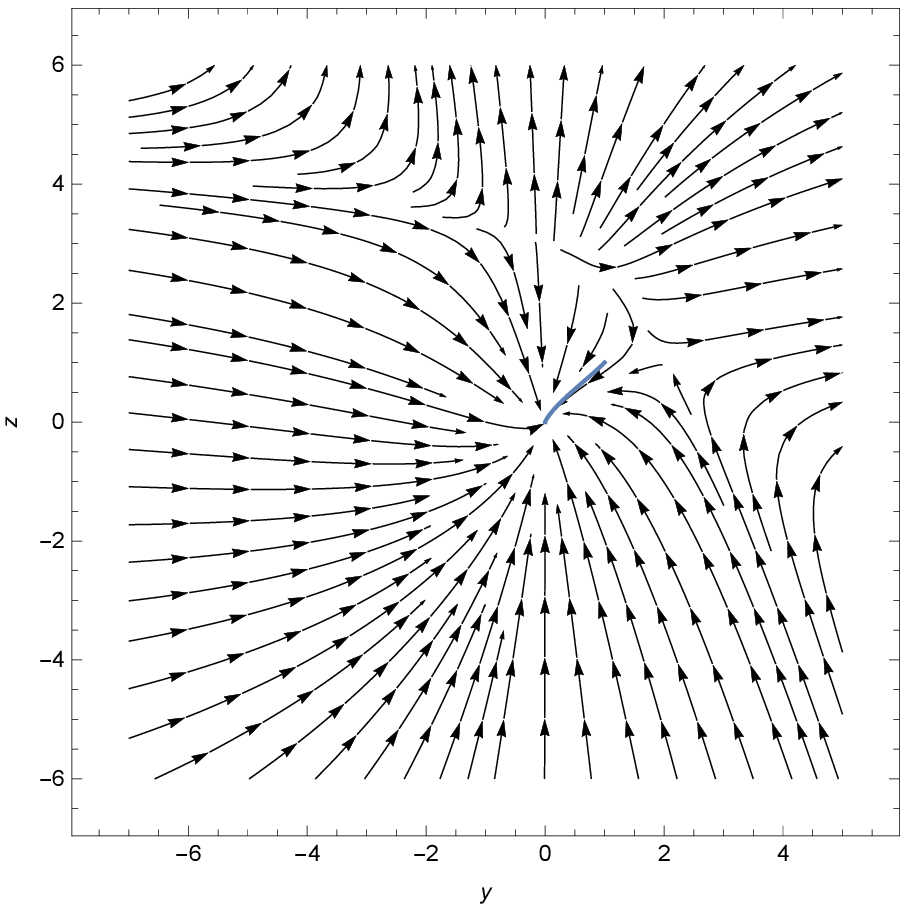}
\includegraphics[height=2in]{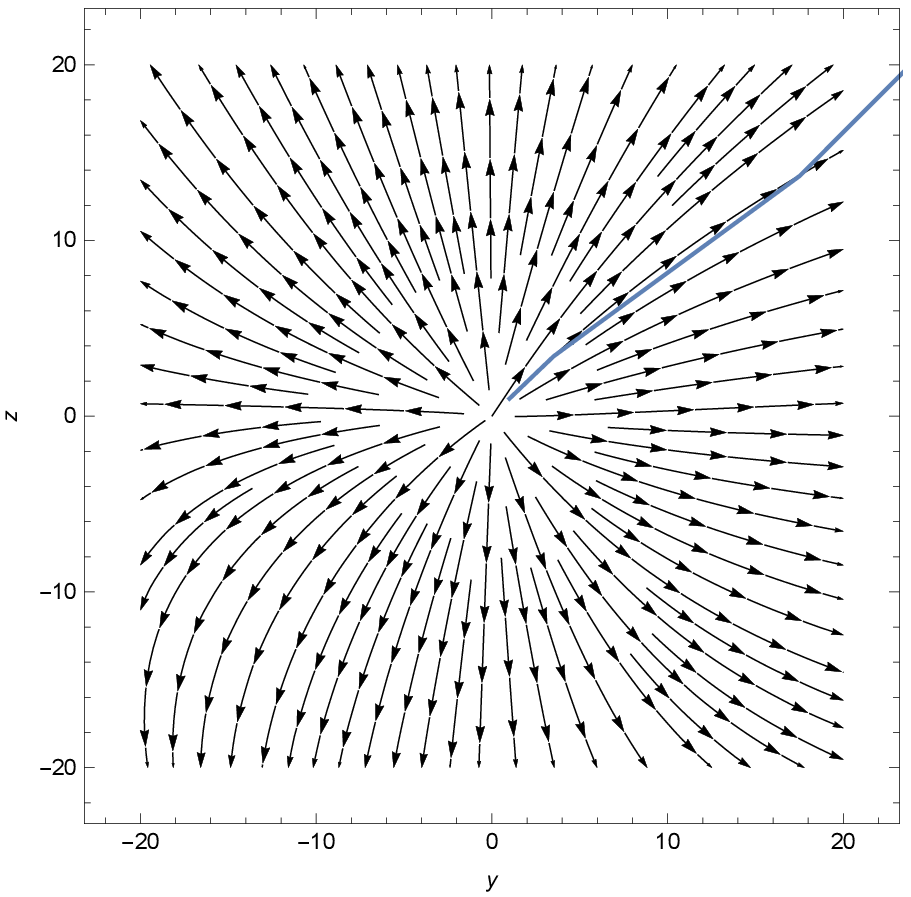}\\
\vspace{1mm}
~~~~~~~~~~~~~~~~~~~~~Fig. 6~~~~~~~~~~~~~~~~~~~~~~~~~~~~~~~~~~~Fig. 7\\
\vspace{5mm}Fig. 6 shows the phase plot of stable attractor $(0,0)$ for analysing stability at infinity for case I when $\gamma=0$, $\alpha_{2}<3$ taking $C_{o}=8\pi G=1$.~~~Fig. 7 shows the phase plot of unstable repeller $(0,0)$ for analysing stability at infinity for case I when $\gamma=0$, $\alpha_{2}>3$ taking $C_{o}=8\pi G=1$.
\hspace{2cm} \vspace{6mm}
\end{figure}

\begin{figure}[H]
\includegraphics[height=2in]{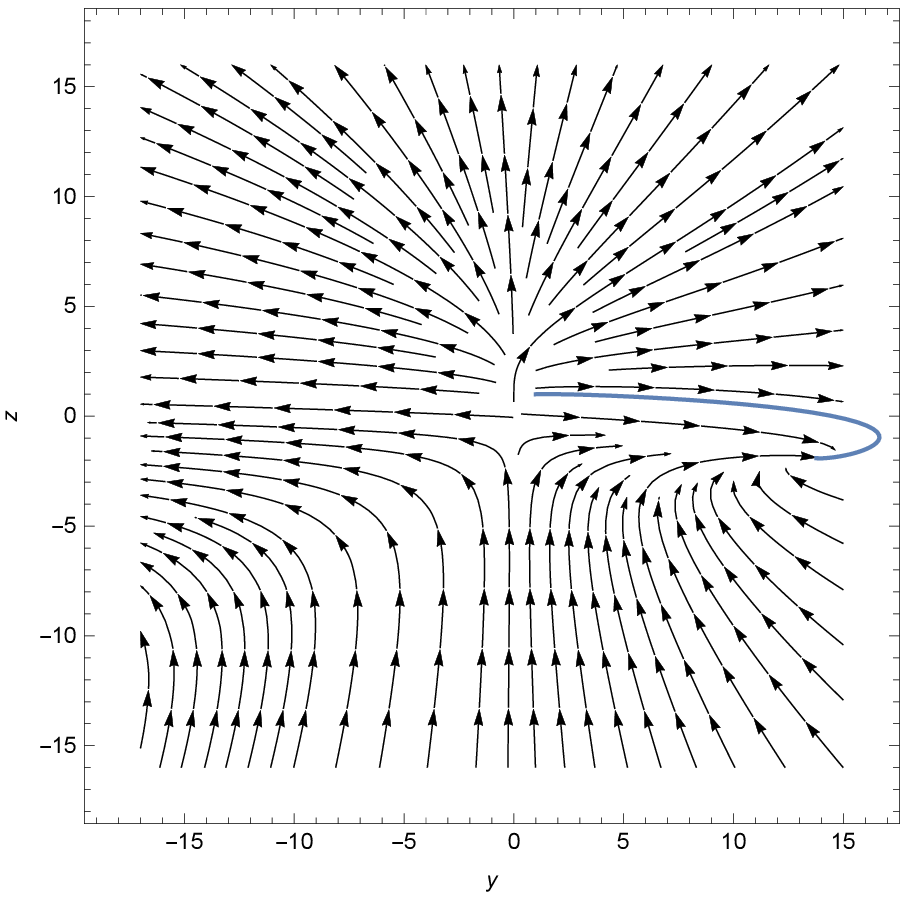}
\includegraphics[height=2in]{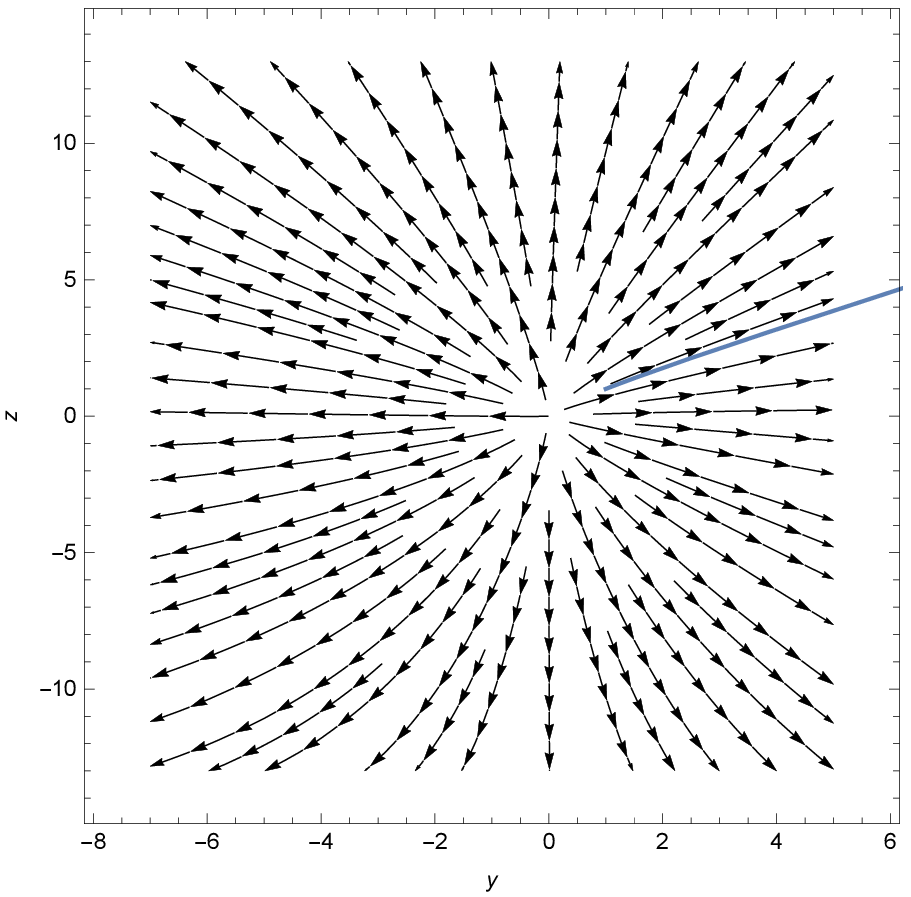}\\
\vspace{1mm}
~~~~~~~~~~~~~~~~~~Fig. 8~~~~~~~~~~~~~~~~~~~~~~~~~~~~~~~~~~~Fig. 9\\
\vspace{5mm}
Fig. 8 shows the phase plot of unstable saddle point $(0,0)$ for analysing stability at infinity for case I when $\gamma=\frac{4}{3}$, $\alpha_{2}<3$ taking $C_{o}=8\pi G=1$.~~~Fig. 9 shows the phase plot of unstable repeller $(0,0)$ for analysing stability at infinity for case I when $\gamma=\frac{4}{3}$, $\alpha_{2}>3$ taking $C_{o}=8\pi G=1$.
\hspace{2cm} \vspace{6mm}
\end{figure}

 {\bf  Case II- Dynamical system analysis for $\dot{G}\neq 0$ and $\rho_{\Lambda}$=constant}\renewcommand {\theequation}{\arabic{equation}}\\

 Let's rewrite the General Relativity field equations \eqref{EFE} as follows:
 \begin{equation*}
   G_{\mu\nu}-g_{\mu\nu}\Lambda=8\pi G,
 \end{equation*}
where $G_{\mu\nu}=R_{\mu\nu}-\frac{1}{2}g_{\mu\nu}R$ denotes the Einstein tensor.

With general Bianchi identity $\nabla^{\mu}G_{\mu\nu}=0$, the above field equation gives the following relation:
\begin{equation*}
  \nabla^{\mu}(\textbf{T}_{\mu\nu})=\nabla^{\mu}[G(T_{\mu\nu}+g_{\mu\nu}\rho_{\Lambda})]=0.
\end{equation*}
 This implies that the local conservation law takes the following form which we named it mixed local conservation law:
 \begin{equation}\label{G}
   \frac{d}{dt}[G(\rho_{m}+\rho_{\Lambda})]+3GH(\rho_{m}+p_{m})=0.
 \end{equation}
 If we assume that $\dot{G}\neq 0 $ and $\rho_{\Lambda}=$constant, then the above relation leads to the following equation which indicates a non-conservation of matter as $G$ does not remain constant here:
 \begin{equation}\label{FE2}
   \dot{G}(\rho_{m}+\rho_{\Lambda})+G[\dot{\rho_{m}}+3H(\rho_{m}+p_{m})]=0.
 \end{equation}

 But if we take $\dot{G}\neq 0$ as well as $\dot{\rho_{\Lambda}} \neq 0$ assuming the standard local covariant conservation of matter-radiation \eqref{BCE0}, \eqref{G} leads to the following equation:
 \begin{equation}\label{FE3}
   (\rho_{m}+\rho_{\Lambda})\dot{G}+ G\dot{\rho_{\Lambda}}=0.
 \end{equation}
Since we are inclined to qualitative study of the dynamics of the Universe, we set up a dynamical system for case-II by introducing new variables: $x=\frac{8\pi G}{3H^{2}}$, $y=\rho_{m}$.

With these new variables the field equations can be rewritten as

\begin{eqnarray}\label{Rel}
  && ~~~ 8\pi G \rho_{tt}\equiv8\pi G \rho_{m} + \Lambda = 3H^{2} \nonumber\\
  &&\Rightarrow 8\pi G (\rho_{m}+\rho_{\Lambda}) = 3H^{2} \nonumber\\
  &&\Rightarrow \frac{8\pi G}{3H^{2}}(\rho_{m}+\rho_{\Lambda}) =1 \nonumber\\
  &&\Rightarrow x(y+\rho_{\Lambda}) = 1 \nonumber\\
  &&\Rightarrow \frac{1}{x} =y+\rho_{\Lambda}.
\end{eqnarray}

Again using the Taylor series form of $\Lambda(H)$ in the field equation $8\pi G \rho_{m} + \Lambda=3H^{2}$, we get
\begin{eqnarray}\label{Lo}
  && ~~~~ 8\pi G \rho_{m} + \Lambda = 3H^{2} \nonumber\\
  &&\Rightarrow 8\pi G \rho_{m}+ \Lambda_{o}+\alpha_{2}H^{2} = 3H^{2} \nonumber \\
  &&\Rightarrow \frac{8\pi G\rho_{m}}{3H^{2}}+\frac{\Lambda_{o}}{3H^{2}}+\frac{(\alpha_{2}-3)}{3} =0 \nonumber \\
  &&\Rightarrow \frac{\Lambda_{o}}{3H^{2}}=\frac{(3-\alpha_{2})}{3}-xy.
\end{eqnarray}

Now the dynamical system is represented by the following system of ordinary differential equations:

\begin{eqnarray}
  x' &=& \frac{dx}{dt}\frac{dt}{d\Theta} \nonumber\\
   &=& \frac{8\pi\dot{G}}{3H^{3}}-\frac{2\dot{H}(8\pi G)}{3H^{4}}.
\end{eqnarray}
Using the expression of $\dot{G}$,$\dot{H}$ and $\frac{\Lambda_{o}}{3H^{2}}$ we have found above, we get
\begin{eqnarray}\label{DS2x}
 x'&=& \frac{-x\Lambda_{o}}{H^{2}}-x(\alpha_{2}-3)+3x^{2}(\gamma-1)y, \nonumber  \\
   &=& 3\gamma x^{2}y.
\end{eqnarray}
and
\begin{eqnarray}\label{DS2y}
  y' &=& \frac{dy}{d\Theta}\frac{d\Theta}{dt}, \nonumber\\
   &=& -3\gamma y.
\end{eqnarray}

In order to find the fixed points we equate $x'=0$ and $y'=0$. If $x'=0$, then either $y=0$ or $\gamma=0$ as $x\neq 0$ otherwise if $x=0$, then \eqref{Rel} will be violated. Again if $\gamma=0$ is considered then we get $y=b$ where $b$ is a real constant and $x=a$ where $a,b \in \mathbb{R}$ satisfies $a(b+\rho_{\Lambda})=1$. So the first fixed point we have obtained here is $P=(a,b)$ where $ a(b+\rho_{\Lambda})=1; a,b \in \mathbb{R}$. Now consider $y=0$ when $\gamma \neq 0$ then $x=\frac{1}{\rho_{\Lambda}}$, that is, $Q=(\frac{1}{\rho_{\Lambda}},0)$ is the second fixed point. In studying the stability of the fixed points, Jacobian matrix of the system plays a leading role. The Jacobian matrix $J_{2}$ of the system is as follows:\\

\begin{center}
$J_{2}=\left(
   \begin{array}{cc}
     6\gamma xy & 3\gamma x^{2} \\
     0 & -3\gamma \\
   \end{array}
 \right)
$.
\end{center}

At the fixed points $P$, $Q$, $J_{2}$ takes the following form respectively:

\begin{center}
 $J_{P}=\left(
   \begin{array}{cc}
     6\gamma ab & 3\gamma (\frac{1}{b+\rho_{\Lambda}})^{2} \\
     0 & 0 \\
   \end{array}
 \right)
$.
\end{center}

Since $P$ is obtained when $\gamma=0$, $J_{P}$ becomes a null matrix and hence the eigenvalues of $J_{P}$ are $m_{1}=0$, $m_{2}=0$. The eigenvalues of $J_{Q}$ are $m_{3}=0$, $m_{4}=-3\gamma$. We see that at least one of the eigenvalues vanish at both the fixed points and hence both $P$ and $Q$ are non-hyperbolic. So we need to use the concept of perturbation function as it is easy to analyse the behaviour of the system from the nature of perturbation function expressed in terms of $\Theta$. As $\Theta$ tends to $\infty$, if the perturbation alone each of the axes grows then the fixed point is unstable whereas if the perturbation along each of the axes decays to zero or evolves to a constant value, then the fixed point is stable. We shall not employ Center manifold theory for two dimensional problems as it is simpler to use the method of perturbation function, but for higher dimensional problems as Center manifold theory is one of the prominent tools to study stability of a system, we have also shown in the later part, namely, Case III of this section how the dynamics of the center manifold determines the dynamics of the entire system.

\textbf{A. } {\bf Stability analysis using the concept of Spectral radius of the Jacobian matrix at the respective fixed points:}\renewcommand {\theequation}{\arabic{equation}}

The spectral radius of $J_{P}$ and $J_{Q}$ are given by

\begin{center}
$\sigma_{P}=0<1$, $\sigma_{Q}=max\{|-3\gamma|,0\}=\bigg\{\begin{array}{cc}
                                                  3\gamma& , \gamma>0, \\
                                                  0 & ,\gamma=0.
                                                \end{array}$.
\end{center}

 Since $\sigma_{P}<1$, all the eigenvalues of $J_{P}$ lie inside a unit disc. So $P$ is stable. When $\gamma>0$, $\sigma_{Q}<1$ if $\gamma<\frac{1}{3}$ and $\sigma_{Q}=1$ if $\gamma=\frac{1}{3}$. So, $Q$ is stable for $0\leq \gamma<\frac{1}{3}$ and we can't say whether $Q$ is stable or not if $\gamma=\frac{1}{3}$. In addition when $\gamma=\frac{1}{3}$ one eigenvalue of $J_{Q}$,namely, $-3\gamma$, has absolute value equal to one the other eigenvalue, that is, zero has absolute value less than one. In this case a bifurcation may occur where a small change in the parameter values of the system leads to a sudden qualitative change in terms of topological behavior of the system. We need to further our study from the concept of perturbations along each axes and study the behaviour of perturbations when $\Theta\rightarrow \infty$.\\

\textbf{B. } {\bf Stability analysis using the concept of Perturbation function:}\renewcommand {\theequation}{\arabic{equation}}\\

 Let $x=x_{P}+\eta_{x}$ and $y=y_{p}+\eta_{y}$, where $x_{P}$, $y_{P}$ are the values of $x,y$ at $P$ and $\eta_{x}, \eta_{y}$ are small perturbations along $x-$axis and $y-$axis respectively. Putting the perturbed value of $x$ and $y$ in the dynamical system equations \eqref{DS2x} and \eqref{DS2y} leads to the following relations:\\
\begin{eqnarray*}
   \eta_{x} &=&c_{1},\\
   \eta_{y} &=&c_{1}e^{-3\gamma \Theta}-b,
 \end{eqnarray*}
where $c_{1}$ is an arbitrary constant. Similarly, at fixed point $Q$ we get\\
\begin{eqnarray*}
  \eta_{x} &=& c_{2}, \\
  \eta_{y} &=& c_{2}e^{-3\gamma \Theta},
\end{eqnarray*}
where $c_{2}$ is an arbitrary constant. \\
As $\Theta$ increases and tends to $\infty$, $\eta_{y}$ for $P$ evolves to a constant value for all $\gamma \in [0,2]$ and $\eta_{y}$ for $Q$ also converges to zero for all $\gamma \in [0,2]$. Since the perturbation along each axis does not grow with the increase in $\Theta$, $P$ is stable for all $\gamma \in [0,2]$, in particular for $\gamma=0$. When $\gamma \neq 0$ $\eta_{y}\rightarrow -b$ as $\Theta \rightarrow \infty$ but if we directly put $\gamma=0$ in the expression of $\eta_{y}$ above, $\eta_{y}$ becomes a constant function, $\eta_{y}=c_{1}-b$. Fig. 10 shows the variation of perturbation along $y-$axis , $\eta_{y}$ with respect to $\Theta$ as $\gamma\rightarrow 0^{+}$ for the fixed point $P$. From Fig. 10 we see that as $\gamma\rightarrow 0$ from the right the curves gradually tends to $\eta_{y}=c_{1}-b$. Fig. 11 shows that $\eta_{y}$ decreases exponentially as $\Theta$ increases and ultimately decays to zero as $\Theta$ tends to $\infty$ for $Q$ for any positive value of $\gamma$. So it is obvious that $\eta_{y}\rightarrow 0$ as $\Theta \rightarrow \infty$ for $\gamma=\frac{4}{3}$ also which is $\frac{1}{3}$ as determined from the concept of spectral radius. So $Q$ is also no doubt stable for all $0<\gamma < \frac{1}{3}$. We have calculated the value of effective equation of state parameter $\omega_{eff}=-1-\gamma xy$ and relative energy density $\Omega_{tt}=\Omega_{m}+\Omega_{\Lambda}$, where $\Omega_{m}=xy$, $\Omega_{\Lambda}=\frac{\Lambda_{o}}{3H^{2}}+\frac{\alpha_{2}}{3}=1-xy$. At both the fixed points $P$ and $Q$, we get $\omega_{eff}=-1$, $\Omega_{tt}=1$ which is in agreement with the observational data in \cite{Planck}. Since $\omega_{eff}$ is found to be negative unity, the presence of the stable fixed point $P$ indicates the presence of negative pressure in the developed cosmological model which contributes to our model with an accelerated expansion phase of the Universe. We tabulated the results in TABLE II:\\

\begin{table}[!h]
\caption{Table for case II ($\dot{G}\neq 0$, $\rho_{\Lambda}$=constant)}
\label{table: Table1 }
\resizebox{\columnwidth}{!}{%
\begin{tabular}{|c|c|c|c|c|c|c|c|}
\hline
Fixed points & x & y & Type of fixed point & Eigen Values & $\omega_{eff}$ & $\Omega_{tt}$  & Behavior \\
\hline
$P$ & a, & b & non-hyperbolic& 0 , 0 & -1 & 1 & stable for $\gamma=0$  \\
 & & & & & & & \\
 & where $a(b+\rho_{\Lambda})=1 $ & & & & & & \\
 & & & & & & & \\
 & & & & & & & \\
$Q$ & $\frac{1}{\rho_{\Lambda}}$ & 0 & non-hyperbolic & 0, $-3\gamma$ &-1 & 1 & stable for $0\leq \gamma<\frac{1}{3}$ ,\\
 & & & & & & &\\
\hline
\end{tabular}%
}
\end{table}

\begin{figure}[H]
\includegraphics[height=1.6in]{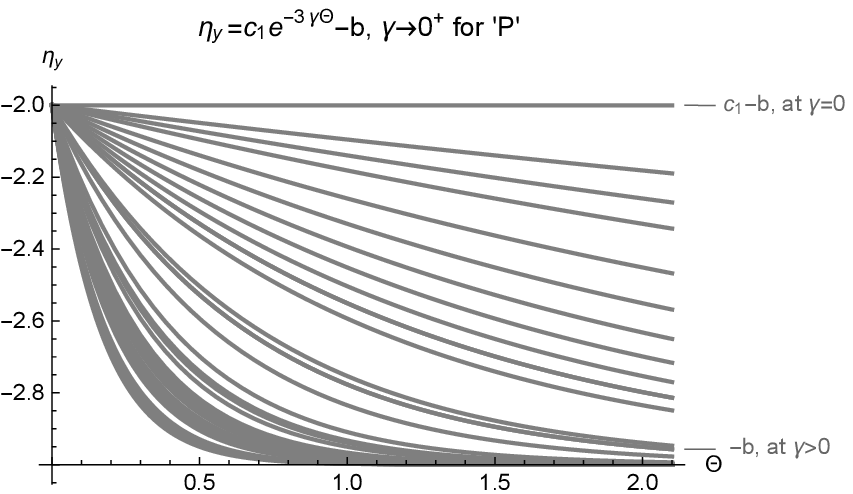}
\includegraphics[height=1.6in]{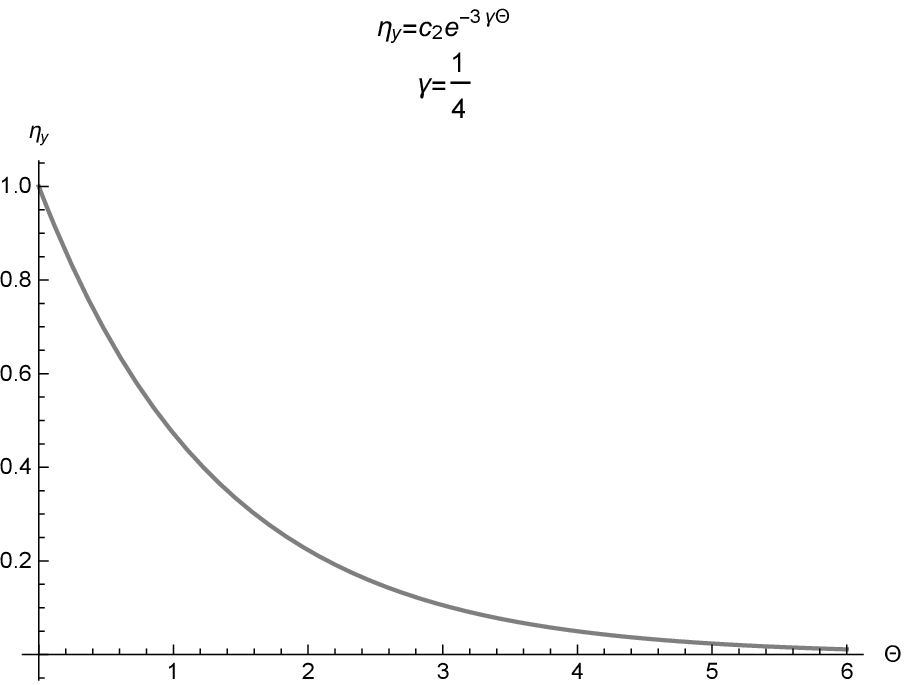}\\
\vspace{1mm}
~~~~~~~~~~~~~~~~~~~Fig. 10~~~~~~~~~~~~~~~~~~~~~~~~~~~~~~~~~~~~~~~~~~~~Fig. 11\\
\vspace{5mm}\\
Fig. 10 shows variation of $\eta_{y}$ with respect to $\Theta$ for fixed point $P$ as $\gamma \rightarrow 0^{+}$.~~~Fig. 11 shows the variation of $\eta_{y}$ with respect to $\Theta$ for $Q$ at $\gamma=\frac{1}{4}<\frac{1}{3}$.
\hspace{2cm} \vspace{6mm}
\end{figure}

 {\bf  Case III- Dynamical system analysis for $\dot{G}\neq 0$ and $\dot{\rho_{\Lambda}}\neq 0$}\renewcommand {\theequation}{\arabic{equation}}\\

In this case both $G$ and $\rho_{\Lambda}$ are no longer constants, that is, $\dot{G}\neq 0$ and $\dot{\rho_{\Lambda}}\neq 0$. The relation in \eqref{G} now becomes

\begin{equation}\label{Gdot}
  \dot{G}(\rho_{m}+\rho_{\Lambda})+G\dot{\rho_{\Lambda}}=0.
\end{equation}

We introduce the following new variables to set up the corresponding dynamical system: $x=\frac{8\pi G}{3H^{2}}$, $y=\rho_{m}$, $z=\rho_{\Lambda}$. We take derivative of the newly introduced variables with respect to logarithmic time, $\Theta$ and obtain the following relations:

\begin{eqnarray*}
  x'&=& \dot{x}\frac{dt}{d\Theta}, \nonumber\\
   &=& \frac{1}{H}\frac{d}{dt}(\frac{8\pi G}{3H^{2}}), \nonumber\\
   &=& \frac{8\pi \dot{G}}{3H^{3}}-\frac{2(8\pi G \dot{H})}{3H^{4}}.
\end{eqnarray*}
Using \eqref{Gdot} in the above equation  and the necessary substitutions we get
\begin{equation}\label{ds3x}
  x'=x^{2}y'+3(3\gamma-1)x^{2}y-(\alpha_{2}-3)x,
\end{equation}

\begin{eqnarray}\label{ds3y}
  y' &=& \dot{\rho_{m}}\frac{dt}{d\Theta}, \nonumber\\
   &=& \frac{1}{H}\dot{\rho_{m}}, \nonumber\\
   &=& \frac{1}{H}(\dot{-\rho_{\Lambda}}-3\gamma H\rho_{m}), \nonumber\\
   &=& -z'-3\gamma y,
\end{eqnarray}

\begin{eqnarray}\label{DS3z}
  z' &=& \dot{\rho_{\Lambda}}\frac{dt}{d\Theta}, \nonumber\\
   &=&(6-16\pi \gamma+(\frac{8\pi}{3}-2)\alpha_{2})y+ \nonumber\\
   & & (6+(\frac{8\pi}{3}-2)\alpha_{2}-\frac{16\pi}{3})z-\frac{3y}{x^{2}}.
\end{eqnarray}
Putting the above expression of $z'$ in \eqref{ds3y}, we get the expression of $y'$ as follows:

\begin{eqnarray}\label{DS3y}
   y' &=& (-6+16\pi \gamma-3\gamma -(\frac{8\pi}{3}-2)\alpha_{2})y \nonumber \\
      &&  -(6+(\frac{8\pi}{3}-2)\alpha_{2})-\frac{16\pi}{3})z+\frac{3y}{x^{2}}.
\end{eqnarray}

Finally putting the value of $y'$ above in \eqref{ds3x}, we get the expression of $x'$ as follows:

\begin{eqnarray}\label{DS3x}
 x' &=&-(\alpha_{2}-3)x+3y+(-9+16\pi\gamma+6\gamma+(2-\frac{8\pi}{3})\alpha_{2})x^{2}y \nonumber\\
    &&  -(6+(\frac{8\pi}{3}-2)\alpha_{2}-\frac{16\pi}{3})x^{2}z.
\end{eqnarray}

The expression of total energy density $\Omega_{tt}$ and effective equation of state $\omega_{eff}$ in terms of the variables $x,y,z$ is as follows:
\begin{eqnarray}\label{omega3}
  \Omega_{tt} &=& xy+\frac{z}{y+z}, \\
  \omega_{eff} &=& \frac{p_{tt}}{\rho_{tt}},
\end{eqnarray}
where $p_{tt}=(\gamma-1)y-z$ and $\rho_{tt}=y+z$. We equate $x'=0$, $y'=0$, $z'=0$ using \eqref{DS3x},\eqref{DS3y} and \eqref{DS3z} to obtain the fixed points. As $y\rightarrow 0$, $z\rightarrow 0$, then since $x,y,z$ holds the relation $\frac{1}{y+z}=x$, $x$ must tend to infinity. If we view from the sequential approach of real analysis, any real sequence of the form $\frac{1}{n}$ converges to zero as $n\rightarrow \infty$ but never equals to zero. For every $\epsilon>0$ there exist a positive integer $m$ such that $|\frac{1}{n}-0|<\epsilon$ for all $n\geq m$, that is, in every neighbourhood of zero there contains infinite members of the sequence $\frac{1}{n}$. Similarly when $n\rightarrow 0$, $\frac{1}{n}\rightarrow \infty$. So as $y\rightarrow 0$, $z\rightarrow 0$ $x$ must tends to infinity. To ensure that the fixed points obtained are physically feasible with the developed system, $\alpha_{2}$ must be equal to 3 and with this consideration we can analyse our fixed points in the finite phase plane. Let us consider $x'=0$, $y'=0$, $z'=0$ at $\alpha_{2}=3$, then as $y\rightarrow 0.0009$, $z\rightarrow 0$, $x$ must also tends to a number, $l=\frac{1}{(0.0009+0)}=1111$. Let this fixed point be denoted by $S=(x\rightarrow l,y\rightarrow0.0009,z\rightarrow0)$.

Stability of the above fixed points is determined by the eigenvalues of the Jacobian matrix $J_{3}$ of the above dynamical system which is obtained as follows:

\begin{equation*}
J_{3} =
\begin{pmatrix}
2(-9+16\pi\gamma+6\gamma+(2-\frac{8\pi}{3})\alpha_{2})xy & 3+(-9+16\pi\gamma+6\gamma & -(6+(\frac{8\pi}{3}-2)\alpha_{2}-\frac{16\pi}{3})x^{2} \\
-\alpha_{2}+3-2(6+(\frac{8\pi}{3}-2)\alpha_{2}-\frac{16\pi}{3})xz & +(2-\frac{8\pi}{3})\alpha_{2})x^{2} &  \\
 &  &  \\
-6\frac{y}{x} & (-6+(16\pi-3)\gamma-(\frac{8\pi}{3}-2)\alpha_{2})  & -(6+(\frac{8\pi}{3}-2)\alpha_{2}-\frac{16\pi}{3}) \\
 & +\frac{3}{x^{2}} &  \\
 &  &  \\
6\frac{y}{x} & (6-16\pi \gamma+(\frac{8\pi}{3}-2)\alpha_{2})-\frac{3}{x^{2}} & (6+(\frac{8\pi}{3}-2)\alpha_{2}-\frac{16\pi}{3})
\end{pmatrix}
\end{equation*}

The above matrix is a $3\times3$ matrix. The eigenvalues of $J_{3}$ at the fixed point determines the stability of the fixed point. At $S$ when $\gamma=0$, $J_{3}$ takes the following form:\\

$J_{3}(S)=\left(
  \begin{array}{ccc}
     -(\alpha_{2}-3) & 3+(-9-6.37\alpha_{2})l^{2} & -(-10.74+6.37\alpha_{2})l^{2} \\
    0 & (-6-6.37\alpha_{2})+\frac{3}{l^{2}}& -(-10.74+6.37\alpha_{2}) \\
    0 & (6+6.37\alpha_{2})-\frac{3}{l^{2}} & (-10.74+6.37\alpha_{2}) \\
  \end{array}
\right)$

The above matrix is a $3\times3$ matrix with eigenvalues $0$, $-16.74$, $-(\alpha_{2}-3)=0$. Since some of the eigenvalues becomes zero, $S$ is a non-hyperbolic fixed point. We analyse stability through perturbation function and center manifold theory as it is a three dimensional problem with the fixed point as non-hyperbolic one and using these methods are more suitable.

\textbf{A. } {\bf  Stability analysis for $S$ using the concept of Perturbation function :}\renewcommand {\theequation}{\arabic{equation}}\\

We perturb the system by a small amount putting $x=x_{F}+\eta_{x}, y=y_{F}+\eta_{y}, z=z_{F}+\eta_{z}$ where $x_{F},y_{F},z_{F}$ represent the values of $x,y,z$ at the fixed point to be analyzed for stability and $\eta_{x},\eta_{y},\eta_{z}$ denote the perturbations along $x,y,z$ axes respectively. With these perturbed values in the dynamical system equations \eqref{DS3x}, \eqref{DS3y} and \eqref{DS3z} and necessary substitutions, we obtain the following perturbations as a function of logarithmic time $\Theta$:

\begin{eqnarray}\label{pf3x}
  \eta_{x} &=&\left\{
                \begin{array}{ll}
                  C_{1}e^{-(\alpha_{2}-3)\Theta}-1, & \hbox{for any $\gamma$;} \\
                   C_{1}-1, & \hbox{for any $\gamma$ and $\alpha_{2}=3$.}
                \end{array}
              \right.
\end{eqnarray}

\begin{eqnarray}\label{pf3y}
\eta_{y}&=& \left\{
                 \begin{array}{ll}
                 C_{2}e^{-(6+6.4\alpha_{2})\Theta}, & \hbox{$\gamma$=0;} \\
                   C_{2}e^{(57-6.4\alpha_{2})\Theta}, & \hbox{$\gamma=\frac{4}{3}$;} \\
                   C_{2}e^{(88.5-6.4\alpha_{2})\Theta}, & \hbox{$\gamma=2$.}
                 \end{array}
               \right.
\end{eqnarray}

\begin{eqnarray}\label{pf3z}
  \eta_{z} &=&\left\{
                \begin{array}{ll}
                  C_{3}e^{(-10.7+6.4\alpha_{2})\Theta}, & \hbox{for any $\gamma$;} \\
                  C_{3}e^{8.5\Theta}, & \hbox{for any $\gamma$ and $\alpha_{2}=3$.}
                \end{array}
              \right.
\end{eqnarray}
where $C_{i}, i\in \kappa$ are arbitrary constants and $\kappa$ is the index set.\\
 Let $\Phi=\{\alpha_{2}: \eta_{x} \rightarrow 0$ or $c$, $\eta_{y}\rightarrow 0$ or $c$, $\eta_{z}\rightarrow 0$ or $c$ as $\Theta \rightarrow \infty$, where $c \in \mathbb{R}$ is any real constant \}. If we consider only the expression of $\eta_{x}$ obtained as a function of $\Theta$ regardless of restricting the value of $\alpha_{2}$, then we can see that when $\Theta\rightarrow \infty$, $\eta_{x}\rightarrow C_{1}-l$ for $\alpha_{2}=3$,  $\eta_{x}\rightarrow -l$ for $\alpha_{2}>3$, $\eta_{y}\rightarrow C_{2}$ for any positive value of $\alpha_{2}$. Similarly it is seen that $\eta_{z}$ exponentially increases for $\alpha_{2}>1.67$. So we fail to obtain such value of $\alpha_{2}$ where all of these $\eta_{x},\eta_{y},\eta_{z}$ decay or evolve to a constant value as $\Theta$ tends to infinity. So $\Phi$ is an empty set. Only when all of these $\eta_{x}$, $\eta_{y}$ and $\eta_{z}$ decay to zero or tends to a constant value when $\Theta\rightarrow \infty$, we can conclude that the fixed point is stable otherwise unstable if at least one of them go on increasing as $\Theta\rightarrow \infty$. For $S$ to be stable $\Phi$ should not be an empty set. Fig. 12, Fig. 13 and Fig. 14 show the perturbation plots for $S$ at $\gamma=0$. From Fig. 12, as $\alpha_{2}\rightarrow 3^{-}$, the slope of the curve gradually decreases and as $\alpha_{2}$ becomes exactly equal to 3, the slope of the curve equals zero and then as $\alpha_{2}$ becomes just greater than 3, $\eta_{x}$ becomes an exponentially decreasing function of $\Theta$. So when $\alpha_{2}>3$ as $\Theta \rightarrow \infty$, $\eta_{x}$ exponentially decreases and evolves to a constant value, namely, $-l$. Fig. 13 shows that $\eta_{y} \rightarrow 0$ as $\Theta \rightarrow \infty$ for $\gamma=0$ and any value of $\alpha_{2}$. But from Fig. 14 it is clear that when $\alpha_{2}\geq 3$, $\eta_{z}$ exponentially increases as $\Theta$ increases and continue to grow as $\Theta \rightarrow \infty$. So $S$ is unstable for any value of $\alpha_{2}$. Hence, $S$ is unstable for $\alpha_{2}=3$ also. In this case III, we have already presumed $\alpha_{2}$ to be equal to 3 in order to ensure that the fixed point $S$ obtained above is physically feasible with respect to the dynamical system we have set up. So using the above arguments we conclude that $S$ is unstable from the side of perturbation function. We will also show the use of Center manifold theory in determining the stability of the fixed point $S$. Center manifold theory is one of the most powerful tools to determine stability for non-hyperbolic fixed points as the nature of orbits on a center manifold reflects the nature of the system in the neighbourhood of the fixed point. To use Center manifold theory we need to transform the dynamical system equations into the standard form to study center manifold theory. We know that $S(x\rightarrow l,y\rightarrow 0.0009,z\rightarrow 0)$ is a non-hyperbolic fixed point. Now using a suitable coordinate transformation we need to transformed the system in the required standard form to apply Center manifold theory for it will not change the nature of the fixed point. We present how to analyse stability using the Center manifold theory in the following section.

\begin{figure}[H]
\includegraphics[height=1.6in]{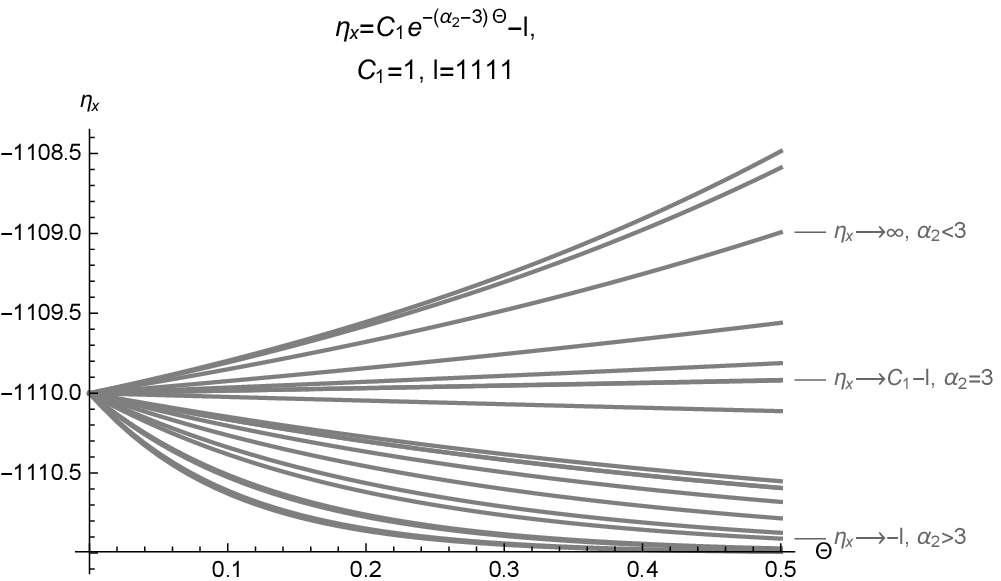}
\includegraphics[height=1.6in]{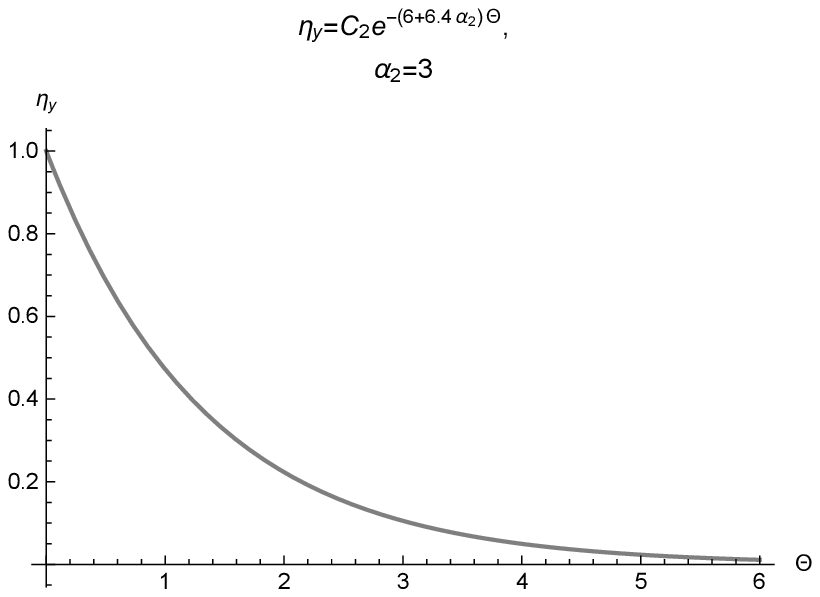}\\
\vspace{1mm}
~~~~~~~~~~~~~~~~Fig. 12~~~~~~~~~~~~~~~~~~~~~~~~~~~~~~~~~~~~~~~~~~~~~~~Fig. 13\\
\vspace{5mm}\\
Fig. 12 shows variation of $\eta_{x}$ with respect to $\Theta$ for $S$.~~~Fig. 13 shows variation of $\eta_{y}$ with respect to $\Theta$ for $S$ at $\gamma=0$
\hspace{2cm} \vspace{6mm}
\end{figure}

 \begin{figure}[H]
\includegraphics[height=1.6in]{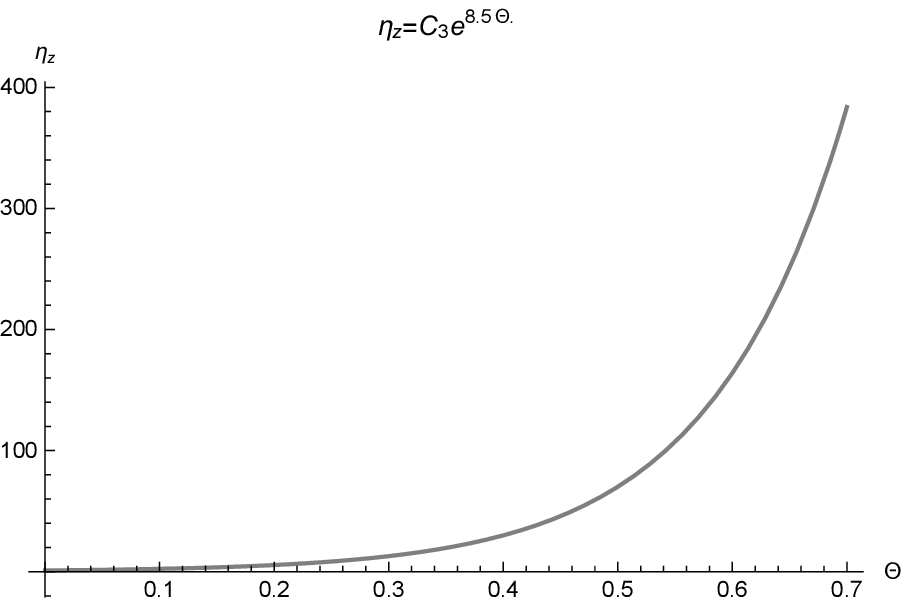}
\includegraphics[height=1.6in]{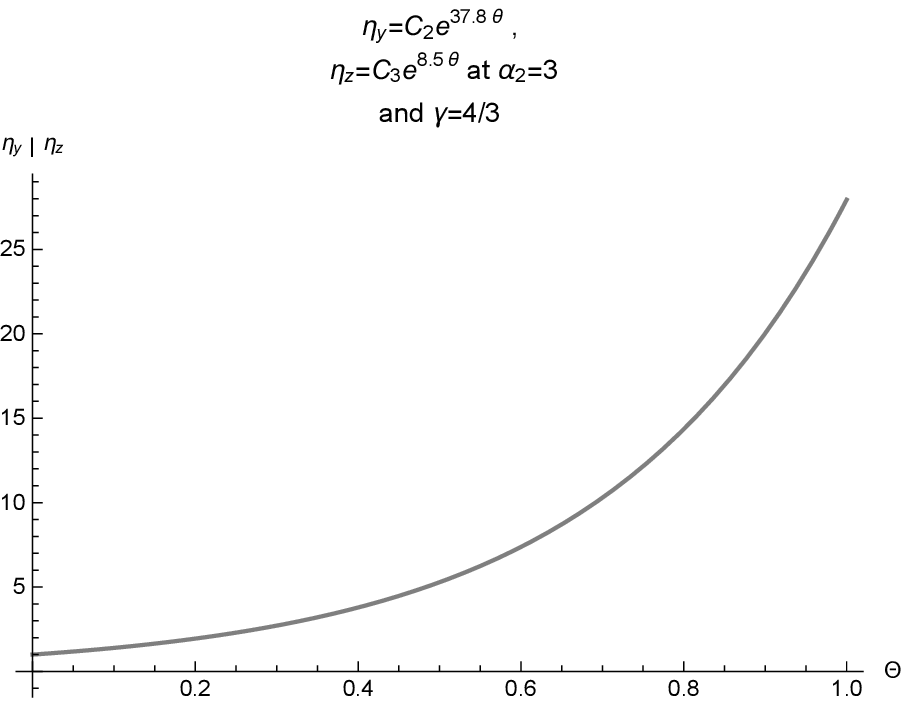}
\vspace{1mm}
~~~~~~~~~~~~~~~~Fig. 14~~~~~~~~~~~~~~~~~~~~~~~~~~~~~~~~~~~~~~~~~~~~~~Fig. 15
\vspace{5mm}\\
Fig. 14 shows the variation of $\eta_{z}$ with respect to $\Theta$ for $S$ at $\gamma=0$.~~~Fig 15 shows variation of $\eta_{y}$ and $\eta_{z}$ at $\gamma=\frac{4}{3}$ and $\alpha_{2}=3$.
\hspace{2cm} \vspace{6mm}
\end{figure}

\begin{figure}[H]
\includegraphics[height=1.6in]{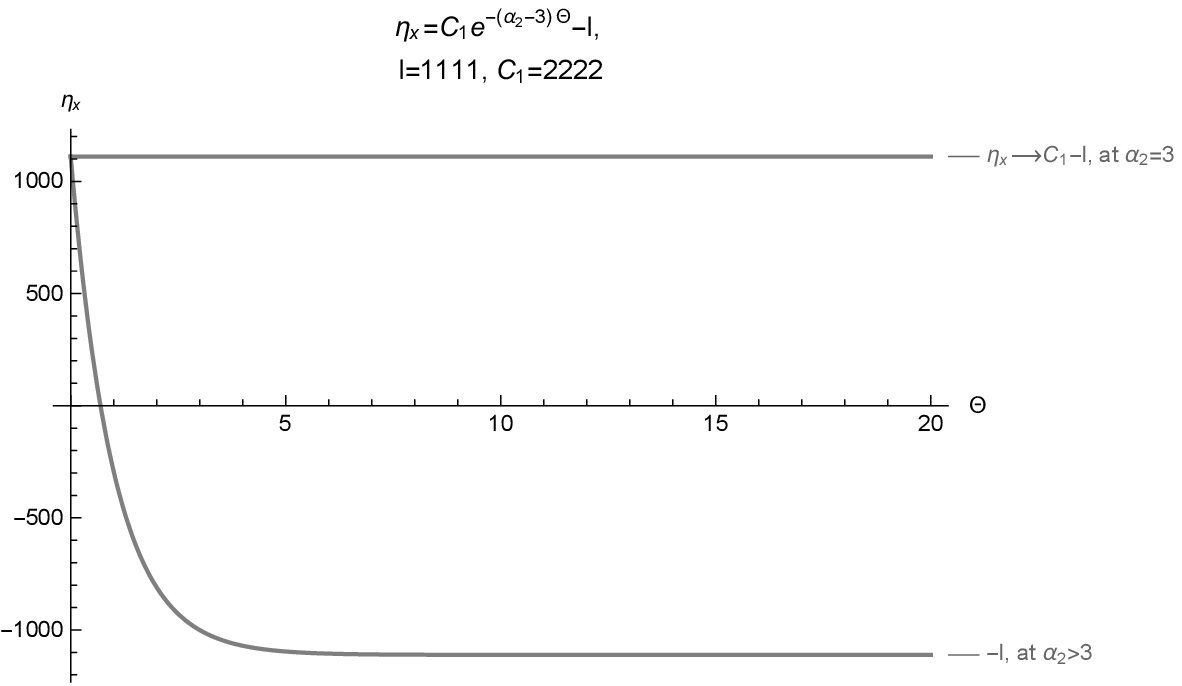}
\includegraphics[height=1.6in]{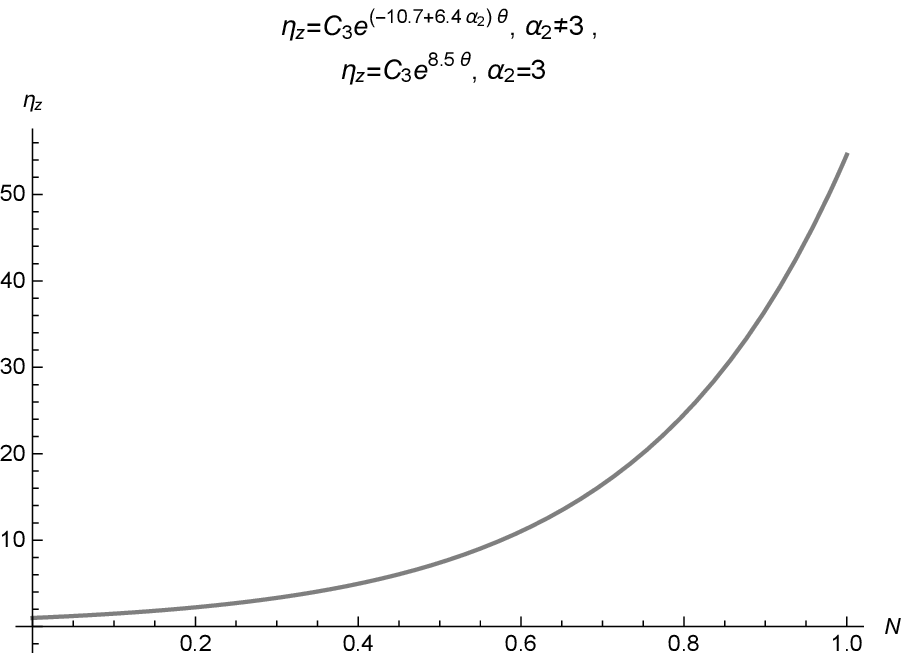}\\
\vspace{1mm}
~~~~~~~~~~~~~~~~~Fig.16~~~~~~~~~~~~~~~~~~~~~~~~~~~~~~~~~~~~~~~~~~~~~~~Fig. 17 \\
\vspace{5mm}\\
 Fig. 16 shows variation of $\eta_{x}$ with respect to $\Theta$ for $S$ when $\gamma=\frac{4}{3}$.~~~Fig. 17 shows the variation of $\eta_{z}$ with respect to $\Theta$ at $S$ at $\gamma=2$.
\hspace{2cm} \vspace{6mm}
\end{figure}

\textbf{B. }{\bf Stability analysis for $S$ using Center Manifold Theory:}\renewcommand {\theequation}{\arabic{equation}}\\

Firstly, we need to transform the dynamical system equations into the form required to use center manifold theory. For this we need to shift the fixed point to origin $(0,0,0)$ by doing suitable coordinate transformation as follows:
\begin{equation*}
  X=x-l, Y=y-0.0009, Z=z;
\end{equation*}
 In terms of this new coordinates our dynamical system equations \eqref{DS3x}, \eqref{DS3y}and \eqref{DS3z} with $\alpha_{2}=3$ can be written as follows:\\
$\left(
  \begin{array}{c}
    X' \\
    Y' \\
    Z' \\
  \end{array}
\right)$=
$\left(
  \begin{array}{ccc}
    -0.05 l & (3-28.11 l^{2}) & -8.37 l^{2} \\
    0 & -25.11 & -8.3 \\
     & 25.11 & 8.3 \\
  \end{array}
\right)$ $\left(
          \begin{array}{c}
            X \\
            Y \\
            Z \\
          \end{array}
        \right)$
+
$\left(
  \begin{array}{c}
    f(X,Y,Z). \\
  g_{1}(X,Y,Z).\\
  g_{2}(X,Y,Z).\\
  \end{array}
\right)$,

where
\begin{eqnarray*}
   && f(X,Y,Z)=-0.025X^{2}-28.11X^{2}Y-8.37X^{2}Z-(56.22 l)XY-(16.74 l)XZ, \\
   && g_{1}(X,Y,Z)=\frac{3(Y+0.0009)}{(X+l)^{2}}, \\
   && g_{2}(X,Y,Z)=-\frac{3(Y+0.0009)}{(X+l)^{2}}.
\end{eqnarray*}

 The Jacobian matrix of the above system at origin is

\begin{center}
$J_{(X=0,Y=0,Z=0)}$=
$\left(
  \begin{array}{ccc}
    -0.05 l & 3-28.11 l^{2} & -8.3 l^{2} \\
    -\frac{0.0054}{4l^{2}} & -25.11 & -8.3 \\
    \frac{0.0054}{4l^{2}} & 25.11 & 8.3 \\
  \end{array}
\right)$
\end{center}

The above Jacobian matrix has zero determinant which means at least one of the eigenvalues has become zero. To find the eigenvalues say $m_{i}$ we solve the characteristic equation $det(J_{X=0,Y=0,Z=0}-mI)=0$ and obtain $m_{1}=0, m_{2}=-0.05 l, m_{3}=-16.81$. The minimal polynomial that annihilates $J_{X=0,Y=0,Z=0}$ is given by $m(m+0.05 l)(m+16.81)$. As the linear factors occur exactly once in the minimal polynomial, $J_{(X=0,Y=0,Z=0)}$ is diagonalisable. To diagonalise $J_{(X=0,Y=0,Z=0)}$ to obtain the required form to use center manifold theory, we need to find the stable subspace $E^{s}$ generated by the eigenbasis associated with the negative eigenvalues, the center subspace $E^{c}$ generated by the eigenbasis associated with the zero eigenvalue of above Jacobian matrix. The eigenspace associated with zero eigenvalue can be found out by solving for $x_{1},x_{2},x_{3}$ in the following matrix equation:

\begin{center}
 $ (J-(0)I_{3\times3})\left(
            \begin{array}{c}
              x_{1} \\
              x_{2} \\
              x_{3} \\
            \end{array}
          \right)=O_{3\times3}$,
 \end{center}
where $I_{3\times3}$ and $O_{3\times3}$ represents the identity matrix and null matrix respectively. Solving the above equations we get the eigenbasis as

\begin{center}
 $E^{c}=\left\{ \left(
            \begin{array}{c}
             -728 l\\
              -1 \\
              1 \\
            \end{array}
          \right) \right\}$
\end{center}

Similarly we find the eigenbasis associated with the eigenvalues $-0.05 l$ and -16.81 so that we can write stable subspace $(E^{s})$ as follows:

 \begin{center}
 $E^{s}=\left\{ \left(
             \begin{array}{c}
               -592 l^{2} \\
               -0.3 \\
               1 \\
             \end{array}
           \right), \left(
                       \begin{array}{c}
                         196 l \\
                         -1 \\
                         1 \\
                       \end{array}
                     \right) \right\}$
 \end{center}

Both $E^{c}$ and $E^{s}$ are the subspaces of $\mathbb{R}\times\mathbb{R} \times \mathbb{R}$. Let us define a matrix $P$ whose column vectors are formed by the above eigenbases as follows:
\begin{center}
$P=\left(
     \begin{array}{ccc}
       -728 l & -592 l^{2} &196 l \\
       -1 & -0.3 & -1 \\
       1 & 1 & 1 \\
     \end{array}
   \right)$
\end{center}

$P$ is a non-singular matrix with $det(P)=-646.8 l$. So $P$ is invertible matrix with $P^{-1}$ as
$P^{-1}=\frac{1}{det(P)}Adj(P)$, where $Adj(P)$ denotes the adjoint of $P$. Therefore

\begin{center}
$P^{-1}=\left(
          \begin{array}{ccc}
            \frac{-0.7}{646.8 l} & -(0.9 l+0.3) & 0.53 \\
            0 & 1.4 & 1.4 \\
            \frac{0.7}{646.8 l} & 0.9 l & 0.9 l \\
          \end{array}
        \right)$.
\end{center}

We again define a new co-ordinate transformation as:
\begin{center}
$P\left(
  \begin{array}{c}
    U \\
    V \\
    W \\
  \end{array}
\right)=\left(
          \begin{array}{c}
            X \\
            Y\\
            Z \\
          \end{array}
        \right)$,
\end{center}

that is,
\begin{center}
$P^{-1}\left(
  \begin{array}{c}
    X \\
    Y \\
    Z \\
  \end{array}
\right)=\left(
          \begin{array}{c}
            U \\
            V\\
            W \\
          \end{array}
        \right)$.
\end{center}

In terms of the new coordinates $U$, $V$, $W$, $X$, $Y$ and $Z$ can be expressed as follows:\\

$X=-728 l U-592 l^{2}V+196 l W$, $Y=-U-0.3V-W$, $Z=U+V+W.$\\

The definition of Center manifold allows us to take $h_{1}$ and $h_{2}$ in Taylor's series form as $V=h_{1}(U)=a_{1}U^{2}+a_{2}U^{3}$ and $W=h_{2}(U)=b_{1}U^{2}+b_{2}U^{3}$ so that $h_{1}(0)=h_{1}(0)=0$ and $Dh_{1}(0)=Dh_{2}(0)=0$, where $D=\frac{d}{dU}$.

We then obtain the required standard form to apply central manifold theory as follows:\\
$
\left(
      \begin{array}{c}
        U' \\
        V' \\
        W' \\
      \end{array}
    \right)=\left(
              \begin{array}{ccc}
                0 & 0 & 0 \\
                0 & -0.05 l & 0 \\
                0 & 0 & -16.81 \\
              \end{array}
            \right)\left(
                     \begin{array}{c}
                       U \\
                       V \\
                       W \\
                     \end{array}
                   \right)+P^{-1}\left(\begin{array}{c}
                             f(U,V,W) \\
                             g_{1}(U,V,W) \\
                             g_{2}(U,V,W)
                           \end{array}
                           \right)$,

where
\begin{eqnarray*}
       f(U,V,W) &=&(-13249 l-40928 l+12186 l^{2})U^{2}+(-21548 l^{3}a_{1}+ \\
                & &7134l^{2}b_{1}+14897850l^{2}-12278a_{1}-40928lb_{1}-33282l^{3}a_{1}+11019l^{2}b_{1}\\
                & &-4435966l^{2}+12186l^{2}a_{1}+12186b_{1}+9910l^{3}a_{1}-3281l^{2}a_{1})U^{3},\\
       g_{1}(U,V,W) &=& -3\frac{-U-0.3V-W}{(-728l U-592l^{2}V+196lW)^{2}}, \\
       g_{2}(U,V,W) &=&3\frac{U+0.3V+W}{(-728l U-592l^{2}V+196lW)^{2}}.
     \end{eqnarray*}
Now computing the above equations we obtain the following relations: \\
\begin{eqnarray}\label{DCM}
  U'&=& \frac{-0.7}{646.8l}\{(-13249 l-40928 l+12186 l^{2})U^{2}+(-21548 l^{3}a_{1} \nonumber \\
    &&  +7134 l^{2}b_{1}+14897850 l^{2}-12278a_{1}-40928lb_{1}-33282l^{3}a_{1}+11019l^{2}b_{1} \nonumber\\
    &&  -4435966l^{2}+12186l^{2}a_{1}+12186b_{1}+9910l^{3}a_{1}-3281l^{2}a_{1})U^{3}\},
\end{eqnarray}

\begin{equation}
  V'=-0.05la_{1}U^{2}-0.05la_{2}U^{3},
\end{equation}

\begin{eqnarray}
  W' &=&-16.81(b_{1}U^{2}+b_{2}U^{3})+\frac{0.7}{646.8l}\{(-13249 l-40928 l+12186 l^{2})U^{2} \nonumber\\
     & & +(-21548 l^{3}a_{1}+7134 l^{2}b_{1}+14897850l^{2}-12278a_{1}-40928lb_{1}-33282l^{3}a_{1} \nonumber\\
     & & +11019l^{2}b_{1}-4435966l^{2}+12186l^{2}a_{1}+12186b_{1}+9910l^{3}a_{1}-3281l^{2}a_{1})U^{3}\}. \nonumber\\
\end{eqnarray}

The dynamics of the center manifold is given by:
\begin{center}
$U'=AU+f(U,h_{1}(U),h_{2}(U))$,
\end{center}
where $A=0$, $V=h_{1}(U),W=h_{2}(U)$.

The tangency condition requires that
\begin{equation}\label{TCV}
 V'-\frac{dh_{1}}{dU}U'=0,
\end{equation}

\begin{equation}\label{TCW}
 W'-\frac{dh_{2}}{dU}U'=0.
\end{equation}

By equating the coefficients of $U^{2}$ and $U^{3}$ in the tangency conditions \eqref{TCV} and \eqref{TCW}, we can find the constants $a_{1},a_{2}$ and $b_{1}, b_{2}$ where we unconsider all the powers of $U$ higher than $U^{3}$. Equating the coefficients of $U^{2}$ and $U^{3}$ in the tangency condition of $V$, we get $a_{1}=a_{2}=0$ and from the tangency conditions of $W$ comparing the coefficient of $U^{2}$, we get\\
\begin{eqnarray*}
&& ~~~ -16.81b_{1}+(\frac{0.7}{646.8})(-54177+12186 l) = 0 \\
&&  \Rightarrow b_{1} = \frac{1}{-16.81}(58.6-13.2 l) \\
&&  \Rightarrow b_{1} = -3.5+0.8 l.
\end{eqnarray*}\\

Since $l$ is a very large number, $b_{1}\sim 0.8l$ and comparing the coefficient of $U^{3}$ we get

\begin{eqnarray*}
&& ~~~ -26.4 b_{1} l = -16.81b_{2}+\frac{0.7}{646.8 l}(18153l^{2}-40928l+12186)b_{1} \\
&&  \Rightarrow 16.81 b_{2} = 36.8l^{2}-35.4l+10.6 \\
&&  \Rightarrow b_{2} = 2l^{2}-2l+0.6.
\end{eqnarray*}
Putting the values of $a_{1},a_{2},b_{1},b_{2}$ in the dynamics of center manifold we get
\begin{equation}\label{DCM3}
  U'=j_{1}U^{2}+j_{2}U^{3}+\mathbb{O}(U^{4}),
\end{equation}
where $j_{1}=(-54177l+12186l^{2})$ and $j_{2}=(14522l^{2}-32742l+9748)(2l^{2}-2l+0.6)$.

Since the first term of $U'$ is in even power of $U$, we deduce instability. If suppose $j_{1}=0$ then we will consider the next term which is in the odd power of $U$. Here if $j_{2}$ is negative then, it is stable otherwise if it is positive then we again achieve instability. But in our case $j_{1}$ never equals zero. So from the side of Center manifold theory we conclude that the fixed point $S$ is unstable.

Now when we take $\gamma=\frac{4}{3}$ then, \eqref{DS3x}, \eqref{DS3y} and $\eqref{DS3z}$ becomes

\begin{eqnarray*}
  x' &=& -(\alpha_{2}-3)x+3y+(6-6.4\alpha_{2})x^{2}y-(-10.7+6.4\alpha_{2})x^{2}z, \\
  y' &=&(57-6.4\alpha_{2})y-(-10.7+6.4\alpha_{2})z+3y(y+z)^{2}, \\
  z' &=&(-61+6.4\alpha_{2})y+(-10.7+6.4\alpha_{2})z-3y(y+z)^{2}.
\end{eqnarray*}

 Now when $\gamma=\frac{4}{3}$ we have the Jacobian matrix at $S$ as follows:\\

\begin{center}
$J_{3}(S)=\left(
  \begin{array}{ccc}
     -(\alpha_{2}-3) & 3-13.2 l^{2} & 0 \\
    0 & (57-6.4\alpha_{2})& -(-10.7+6.4\alpha_{2}) \\
    0 & (-61+6.4\alpha_{2}) & (-10.7+6.37\alpha_{2}) \\
  \end{array}
\right)$
\end{center}

Since we obtain $S$ when $\alpha_{2}=3$, we get the eigenvalues as $m_{1}=(3-\alpha_{2})=0$, $m_{2}=12.8(1.8-\sqrt{3+0.15\alpha_{2}})=-0.7$ and $m_{3}=12.8(1.8+\sqrt{3+0.15\alpha_{2}})=46.8>0$. So for $\gamma=\frac{4}{3}$ at $\alpha_{2}=3$, $S$ becomes non hyperbolic fixed point. We need to analyze stability through perturbation function and Center manifold theory. However stability analysis using Center manifold theory is similar to the above shown. So we will only analyze through perturbation function. From \eqref{pf3x},\eqref{pf3y} and \eqref{pf3z}, we see that for $\alpha_{2}=3$ $\eta_{x}$ tends to a constant, namely, $(C_{1}-l)$ as $\Theta \rightarrow \infty$ but $\eta_{y}$ exponentially increases as $\Theta \rightarrow \infty$. $\eta_{z}$ is also an exponentially increasing function of $\Theta$ and hence it fails to decay or evolve to a constant value as $\Theta \rightarrow
 \infty$. Fig. 15 shows the exponential increasing nature of $\eta_{y}$ and $\eta_{z}$ at $\gamma=\frac{4}{3}$, $\alpha_{2}=3$. Fig. 16 shows the perturbation plot for $\eta_{x}$ as $\Theta$ tends to infinity. So $S$ is unstable at $\alpha_{2}=3$ and $\gamma=\frac{4}{3}$. As the perturbation along each of the axis fail to decay or evolve to a constant value we conclude that $S$ is also unstable for $\gamma=\frac{4}{3}$. For $\gamma=2$ also we can see from \eqref{pf3z} that the perturbation along $z$ axis, namely, $\eta_{z}$ is an exponentially increasing function of $\theta$. So $S$ is unstable for any value of $\alpha_{2}$ for $\gamma=2$ and this is shown in Fig. 17 also. \\

\begin{table}[!h]
\caption{Table for case III ($\dot{G}\neq 0$, $\dot{\rho_{\Lambda}}\neq 0$)}
\label{table: Table3 }
\resizebox{\columnwidth}{!}{%
\begin{tabular}{|c|c|c|c|c|c|}
\hline
Fixed points& Type of fixed point & Eigenvalues & $\omega_{eff}$ & $\Omega_{tt}$  & Behavior \\
\hline
$S$& non-hyperbolic& $-(\alpha_{2}-3)=0$, 0,-16.74&-1 &1 & unstable  \\
$(x\rightarrow l$,$y\rightarrow 0$,$z\rightarrow 0)$ &for $\gamma=0$; & & & & \\
 &non-hyperbolic for & $(3-\alpha_{2})=0$,& & & \\
 & $\gamma=\frac{4}{3}$, &$12.8(1.8-\sqrt{3+0.15\alpha_{2}})$ & & & \\
 &$\alpha_{2}=3$&=-0.7 & & & \\
& &  $12.8(1.8+\sqrt{3+0.15\alpha_{2}})$ & 0 & 1 & unstable;\\
 & &=46.8 & & & \\
 & & & & & \\
& & & & & \\
 &non-hyperbolic & $-(\alpha_{2}-3)=0$, &-0.86 & 1 & \\
 &for $gamma=2$, &$0.096(405.1-\sqrt{157143+4131.84\alpha_{2}})$, & $\approxeq$-1 & &  \\
& &=20.37 & & & unstable \\
 &$\alpha_{2}=3$ & $0.096(405.1+\sqrt{157143+4131.84\alpha_{2}})$ & -1.2 & 1 & \\
& & =57.4&$\approxeq$-1 & & \\
using center & & & & & \\
manifold theory:& & & & & \\
$S$ & non-hyperbolic & $-0.05l$, 0, &-1 & 1 & \\
 $(X \rightarrow 0,Y \rightarrow 0,Z \rightarrow 0)$ & for $\gamma=0$, & & & &unstable. \\
 $X=x-l,Y=y-0.0009$,& $\alpha_{2}=3$ & -16.81 & & & .\\
$Z=z-0$ & & & & & \\
\hline
\end{tabular}%
}
\end{table}

\textbf{C. } {\bf  Stability at infinity and Poincar\'{e} sphere: }\renewcommand {\theequation}{\arabic{equation}}\\

Any polynomial system in rectangular coordinates can be extended to the Poincar\'{e} sphere \cite{Roland}. So the idea of projective geometry done in the case of $\mathbb{R}^{2}$ can be extended to higher dimensions for flows in $\mathbb{R}^{3}$ also. Here, the upper hemisphere of $S^{3}$ can be projected onto $\textbf{R}^{3}$ using the transformation of coordinates given by $x=\frac{X}{W}$, $y=\frac{Y}{W}$, $z=\frac{Z}{W}$ and $X=\frac{x}{\sqrt{1+|x|^{2}}}$, $Y=\frac{y}{\sqrt{1+|x|^{2}}}$, $Z=\frac{z}{\sqrt{1+|x|^{2}}}$ and $W=\frac{1}{\sqrt{1+|x|^{2}}}$ for $\textbf{X}=(X,Y,Z,W)\in S^{3}$ with $|\textbf{X}|=1$ and for $\textbf{x}=(x,y,z)\in \mathbb{R}^{3}$. Now we consider the dynamical system equations \eqref{DS3x}, \eqref{DS3y} and $\eqref{DS3z}$ in the following way:

\begin{equation}\label{Case3polynomial}
  \left.\begin{array}{c}
  x' =P_{1}(x,y,z), \\
  y' =P_{2}(x,y,z), \\
  z' =P_{3}(x,y,z),
\end{array}\right\}
\end{equation}
where
\begin{eqnarray*}
 P_{1}(x,y,z) &=&-(\alpha_{2}-3)x+3y+(-9+16\pi\gamma+6\gamma+(2-\frac{8\pi}{3})\alpha_{2})x^{2}y\\
              &&  -(6+(\frac{8\pi}{3}-2)\alpha_{2}-\frac{16\pi}{3})x^{2}z, \\
 P_{2}(x,y,z)&=& (-6+16\pi \gamma-3\gamma -(\frac{8\pi}{3}-2)\alpha_{2})y-\\
              &&  (6+(\frac{8\pi}{3}-2)\alpha_{2})-\frac{16\pi}{3})z+3y(y+z)^{2},\\
 P_{3}(x,y,z)&=& (6-16\pi \gamma+(\frac{8\pi}{3}-2)\alpha_{2})y+\\
              && (6+(\frac{8\pi}{3}-2)\alpha_{2}-\frac{16\pi}{3})z-3y(y+z)^{2}.
\end{eqnarray*}
We have used the relation $\frac{1}{x}=(y+z)$ in \eqref{DS3y} and $\eqref{DS3z}$ above for our convenience with polynomial functions of $x$, $y$, $z$ with maximum degree 3 on the right side of \eqref{Case3polynomial}. Let us denote the maximum degree terms in $P_{1}$, $P_{2}$ and $P_{3}$ by  $\bar{P}_{1}$, $\bar{P}_{2}$ and $\bar{P}_{3}$ respectively. Then we have,

\begin{equation}
\left.\begin{array}{c}
  \bar{P}_{1}(x,y,z) =(-9+16\pi\gamma+6\gamma+(2-\frac{8\pi}{3})\alpha_{2})x^{2}y \\
   -(6+(\frac{8\pi}{3}-2)\alpha_{2}-\frac{16\pi}{3})x^{2}z,\\
  \bar{P}_{2}(x,y,z) = 3y(y+z)^{2}, \\
  \bar{P}_{3}(x,y,z) =-3y(y+z)^{2}.
\end{array}\right\}
\end{equation}

In terms of $X$, $Y$, $Z$ we express the above polynomials as follows:

\begin{equation}
\left.\begin{array}{c}
  \bar{P}_{1}(X,Y,Z) =(-9+16\pi\gamma+6\gamma+(2-\frac{8\pi}{3})\alpha_{2})X^{2}YW^{-3} \\
                -(6+(\frac{8\pi}{3}-2)\alpha_{2}-\frac{16\pi}{3})X^{2}ZW^{-3},  \\
  \bar{P}_{2}(X,Y,Z) = 3Y(Y+Z)^{2}W^{-3}, \\
  \bar{P}_{3}(X,Y,Z) =-3Y(Y+Z)^{2}W^{-3}.
\end{array}\right\}
\end{equation}

Theorem \ref{Theorem3} determines the location of the critical points at infinity for the above polynomial system by considering the following equations:

\begin{eqnarray}\label{1}
&&  ~~~ X \bar{P}_{2}(X,Y,Z)-Y\bar{P}_{1}(X,Y,Z) = 0 \nonumber\\
&&  \Rightarrow 3(Y+Z)^{2}-(-9+56.24\gamma-6.4\alpha_{2})XY+(6.4\alpha_{2}-10.74)XZ=0. \nonumber\\
\end{eqnarray}

\begin{eqnarray}\label{2}
&& ~~~ X \bar{P}_{3}(X,Y,Z)-Z\bar{P}_{1}(X,Y,Z) = 0\nonumber\\
&&  \Rightarrow -3Y(Y+Z)^{2}-(-9+56.24\gamma-6.4\alpha_{2})XYZ+(6.4\alpha_{2}-10.74)XZ^{2}=0. \nonumber\\
\end{eqnarray}

\begin{eqnarray}\label{3}
&&  ~~~ Y \bar{P}_{3}(X,Y,Z)-Z\bar{P}_{2}(X,Y,Z) = 0 \nonumber\\
&&  \Rightarrow 3Y(Y+Z)^{2}(-Y-Z)= 0 \nonumber \\
&&  \Rightarrow \mbox{either}~~ Y=0~~\mbox{or}~~Y=-Z.
\end{eqnarray}

If $Y=0$ then from \eqref{2} we get either $X=0$ or $Z=0$. If $Y=0$ and $X=0$ is considered then from \eqref{1} we see that $Z=0$. But since $X^{2}+Y^{2}+Z^{2}=1$ must hold, the condition $X=0$ is neglected. If $Z=0$ when $Y=0$ in $X^{2}+Y^{2}+Z^{2}=1$, we get $X=\pm1$. So $(\pm1,0,0,0)$ is a fixed point. Also from \eqref{3} if we consider $Y=-Z$, then from \eqref{2} we get either $X=0$ or $Z=0$. If $X=0$ then $X^{2}+Y^{2}+Z^{2}=1$ does not hold. So when $Y=-Z$ and $Z=0$ then $(\pm1,0,0,0)$ is a fixed point. Using Theorem 4 we obtain that the flow defined by \eqref{Case3polynomial} in a neighbourhood of $(\pm1,0,0,0)\in S^{3}$ is topologically equivalent to the flow defined by the following system:

\begin{equation}\label{4}
\left.\begin{array}{c}
\pm y' = (\alpha_{2}-3)yw^{2}+3y^{2}w^{2}-(-6+(16\pi-3)\gamma-6.4\alpha_{2})yw^{2}\\
 +(6+6.4\alpha_{2}-\frac{16\pi}{3})zw^{2}+(-9+56.24\gamma-6.4\alpha_{2})y^{2}\\
 -(-10.74+6.4\alpha_{2})zy-3y(y+z)^{2},\\
\pm z' = (\alpha_{2}-3)zw^{2}+3yzw^{2}+(-9+56.24\gamma-6.4\alpha_{2})yz\\
  -(-10.74+6.4\alpha_{2})z^{2}-(6-16\pi\gamma+6.4\alpha_{2})yw^{2} \\
  -(-10.74+6.4\alpha_{2})zw^{2}+3y(y+z)^{2},\\
\pm w' = (\alpha_{2}-3)w^{3}+3yw^{3}+(-9+56.24\gamma-6.4\alpha_{2})yw-(-10.74+6.4\alpha_{2})zw.\\
\end{array}\right\}
\end{equation}

The Jacobian matrix of the above system at the fixed point $(0,0,0)$ is a null matrix which has all the eigenvalues as zero. So it is a non-hyperbolic fixed point. We will analyse the stability by finding perturbation functions along each of the axis as a function of logarithmic time $\Theta$ by perturbing the system \eqref{4} by a small amount. If the system comes back to the fixed point following the perturbation then the system is stable otherwise if the perturbation grows to make the system moves away from the fixed point then the system is unstable. Nandan Roy and Narayan Banerjee \cite{Nandan} has also used the concept of perturbation function to analyse stability for non-hyperbolic fixed points for three dimensional systems where linear stability fails. Now firstly consider the expression of \eqref{4} corresponding to $+y$, $+z$ and $+w$ respectively. Then we perturbed our system \eqref{4} by taking $y=\eta_{y}$, $z=\eta_{z}$ and $w=\eta_{w}$.

\begin{eqnarray*}
&& ~~~ \frac{d\eta_{y}}{d\Theta} = (-9+56.24\gamma-6.4\alpha_{2})\eta_{y}^{2}-3\eta_{y}^{3}, \\
&& \Rightarrow  d\Theta = \frac{d\eta_{y}}{(-9+56.24\gamma-6.4\alpha_{2})\eta_{y}^{2}-3\eta_{y}^{3}}, \\
&& \Rightarrow  d\Theta =\frac{Ad\eta_{y}}{\eta_{y}}+\frac{Bd\eta_{y}}{\eta_{y}^{2}}+\\
&& ~~~~~~~~~~~~~ \frac{Cd\eta_{y}}{(-9+56.24\gamma-6.4\alpha_{2})-3\eta_{y}},
\end{eqnarray*}
where
\begin{eqnarray*}
   && A=\frac{3}{(-9+56.24\gamma-6.4\alpha_{2})^{2}}, \\
   && B=\frac{1}{(-9+56.24\gamma-6.4\alpha_{2})}, \\
   && C=\frac{9}{(-9+56.24\gamma-6.4\alpha_{2})^{2}}.
\end{eqnarray*}

Integrating both sides of the above differential equation, we get

\begin{equation}\label{5}
 \Theta=f(\eta_{y})=\frac{C}{3} \ln(\frac{\eta_{y}}{(k-3\eta_{y})})-\frac{B}{\eta_{y}},
\end{equation}
where $k=(-9+56.24\gamma-6.4\alpha_{2})$.

The domain of definition $D_{\Theta}$ of the above function at $\gamma=0$ is

\begin{center}
  $D_{\Theta_{0}}=(-\infty,\frac{k}{3})$; $k=-9-6.4\alpha_{2}$, $\alpha_{2} \in \mathbb{R}^{+}$.
\end{center}

The domain of definition $D_{\Theta}$ of the above function at  $\gamma=\frac{4}{3}$ and $\gamma=2$ respectively are as follows:

\begin{equation*}
  D_{\Theta_{\frac{4}{3}}} =\left\{
                              \begin{array}{ll}
                               (0,\frac{k}{3}), & \hbox{$\alpha_{2}<10.29, k=65.9-6.4\alpha_{2}>0$;}\\
                             (-\infty,\frac{k}{3}), & \hbox{$\alpha_{2}>10.29, k=65.9-6.4\alpha_{2}<0$}
                              \end{array}
                            \right.
\end{equation*}

\begin{equation*}
  D_{\Theta_{2}} =\left\{
                              \begin{array}{ll}
                                (0,\frac{k}{3}), & \hbox{$\alpha_{2}<16.15,k=103.4-6.4\alpha_{2}>0$;}\\
                              (-\infty,\frac{k}{3}), & \hbox{$\alpha_{2}>16.15, k=103.4-6.4\alpha_{2}<0$}
                              \end{array}
                            \right.
\end{equation*}

With the above domain and the choice of $+y$ on the left side of \eqref{4}, we cannot analyse our system for $\Theta\rightarrow \infty$ as $\Theta$ becomes bounded above and unbounded below as $\eta_{y}$ tends to 0, that is, when $\Theta \rightarrow -\infty$, $\eta_{y}\rightarrow 0$. Since we want to analyse the late time behaviour of the Universe as logarithmic time $\Theta \rightarrow \infty$ we only  consider the expressions of \eqref{4} corresponding to $-y$, $-z$ and $-w$ on the left sides of \eqref{4} as follows:

\begin{eqnarray*}
   -y'&=&(\alpha_{2}-3)yw^{2}+3y^{2}w^{2}-(-6+(16\pi-3)\gamma-6.4\alpha_{2})yw^{2}+\nonumber\\
 & &(6+6.4\alpha_{2}-\frac{16\pi}{3})zw^{2}+(-9+56.24\gamma-6.4\alpha_{2})y^{2}-\nonumber \\
 & &(-10.74+6.4\alpha_{2})zy-3y(y+z)^{2},
\end{eqnarray*}

\begin{eqnarray*}
  -z' &=& (\alpha_{2}-3)zw^{2}+3yzw^{2}+(-9+56.24\gamma-6.4\alpha_{2})yz-\nonumber\\
  & &(-10.74+6.4\alpha_{2})z^{2}-(6-16\pi\gamma+6.4\alpha_{2})yw^{2}-\nonumber \\
 & & (-10.74+6.4\alpha_{2})zw^{2}+3y(y+z)^{2},
\end{eqnarray*}

\begin{equation*}
 -w' =(\alpha_{2}-3)w^{3}+3yw^{3}+(-9+56.24\gamma-6.4\alpha_{2})yw-(-10.74+6.4\alpha_{2})zw.
\end{equation*}
With this consideration we get the expression of $\Theta$ as a function of $\eta_{y}$ as follows:
\begin{center}
$\Theta=f((\eta_{y})=\frac{C}{3}\ln \frac{k-3\eta_{y}}{\eta_{y}}+\frac{B}{\eta_{y}}$.
\end{center}

When $\Theta\rightarrow \infty$, $f(\eta_{y}) \rightarrow \infty$ which implies $\eta_{y}\rightarrow 0$. So as $\Theta \rightarrow \infty$ the perturbation along $y-$ axis decays to zero. For analysing the perturbation along $z$ and $w$ axes we consider the expression for $+z$ and $+w$ from \eqref{4} and find out the expression of $\eta_{z}$ and $\eta_{w}$ as follows:

\begin{eqnarray*}
  \eta_{z} &=& \frac{1}{(-10.74+6.4\alpha_{2})\Theta+ c_{1}} ; \alpha_{2}\neq 1.6, \\
  \eta_{w} &=& \pm\frac{1}{\sqrt{2(3-\alpha_{2})\Theta+2c_{2}}},
\end{eqnarray*}
 where $c_{1}$ and $c_{2}$ are arbitrary constants of integration. As $\Theta$ tends to infinity both $\eta_{z}$ and $\eta_{w}$ tends to zero. Fig. 18, Fig. 19 and Fig. 20 show the projection of perturbation along $y$, $z$ and $w$ axes respectively for system \eqref{4}. Since all of $\eta_{y}$, $\eta_{z}$ and $\eta_{w}$ decays to zero as $\Theta$ tends to infinity, we conclude that the fixed point $(\pm1,0,0,0)$ is a stable critical point.

 \begin{figure}[H]
\includegraphics[height=1.8in]{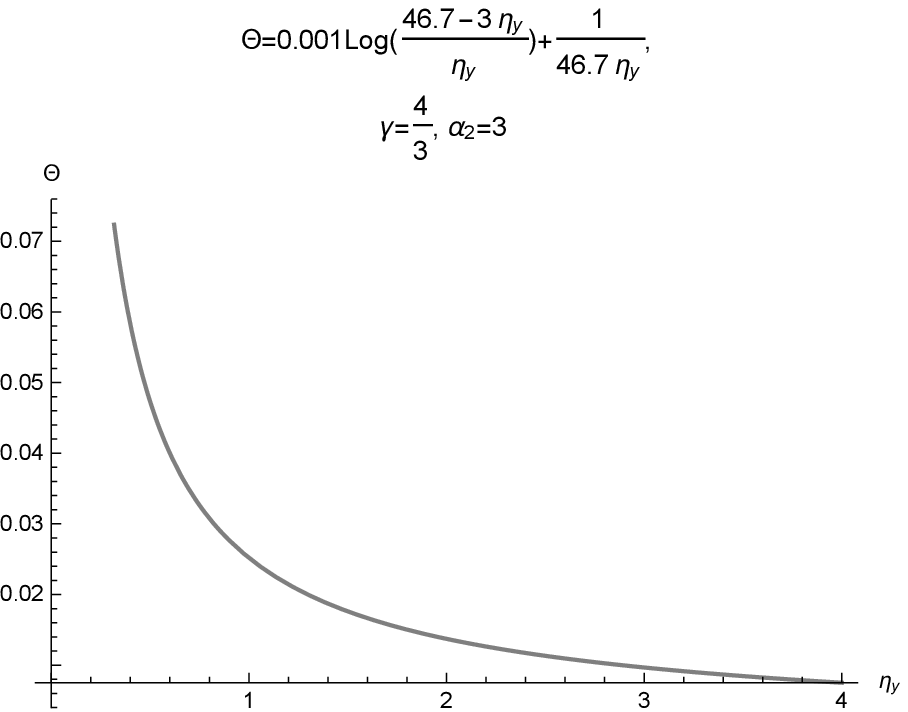}
\includegraphics[height=1.8in]{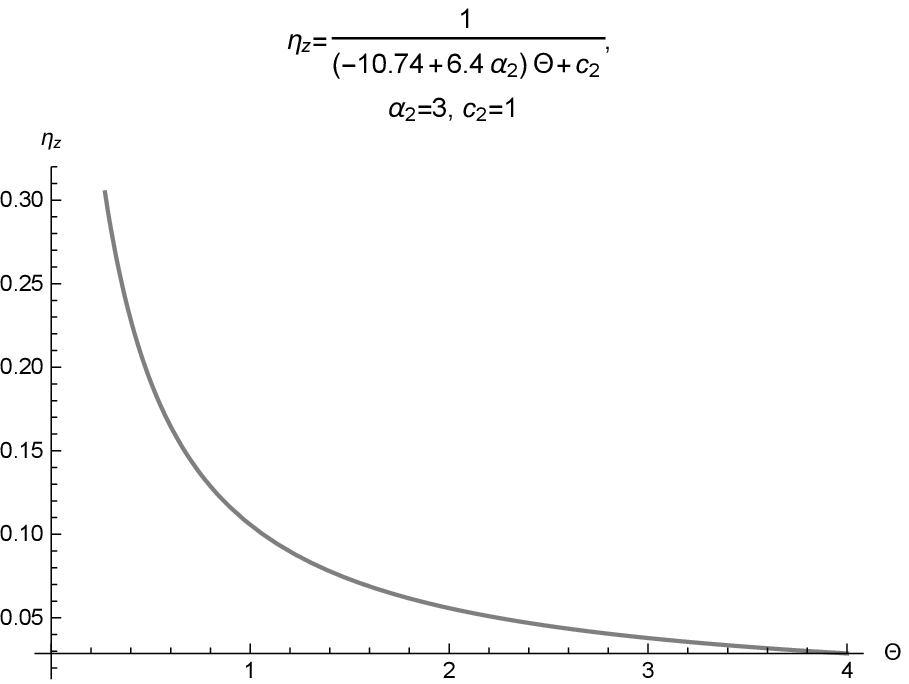}\\
\vspace{1mm}
~~~~~~~~~~~~~~~~~~~Fig. 18~~~~~~~~~~~~~~~~~~~~~~~~~~~~~~~~~~~~~~~~~~~~Fig. 19\\
\vspace{5mm}\\
Fig. 18 shows the variation of $\Theta$ with respect to $\eta_{y}$ for analysing stability at infinity for case III.~~~~Fig. 19 shows the variation of $\eta_{z}$ with respect to $\Theta$ for analysing stability at infinity for case III.
\hspace{2cm} \vspace{6mm}
\end{figure}

\begin{figure}[H]
\includegraphics[height=1.8in]{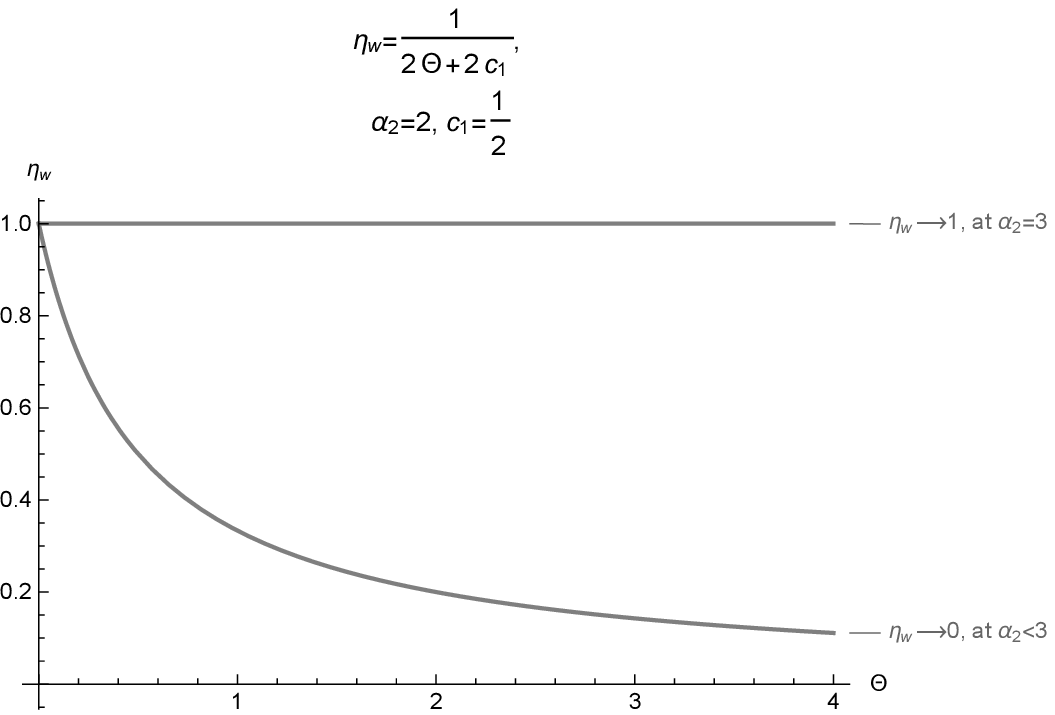}\\
\vspace{1mm}
~~~~~~~~~~~~~~~Fig. 20\\
\vspace{5mm}\\
Fig. 20 shows the variation of $\eta_{w}$ with respect to $\Theta$ for analysing stability at infinity for case III.
\hspace{2cm} \vspace{6mm}
\end{figure}

\section{Conclusion}

In this work we have presented a dynamical system perspective of cosmological models with FLRW metric in the presence of a time varying cosmological constant term which is expressed in Taylor series form of $H$ where we have analyzed for different possibilities of varying $G$ and $\rho_{\Lambda}$. We depict the stability analysis through different approaches by using the concept of spectral radius, perturbation function along each axis and Center manifold theory along with their geometrical analysis. Both analytical and geometrical findings strongly support the fact that the Universe is in the accelerated expansion phase and will continue to expand in the late time also. In case I of section 3 we have shown the model when both $G$ and $\rho_{\Lambda}$ are taken as constants where it becomes similar to the standard $\Lambda CDM$ model. In this section we represent the time varying cosmological constant model as a two dimensional dynamical system having three fixed points $F_{1}$, $F_{2}$, $F_{3}$. The hyperbolic fixed point $F_{1}$ has all its eigenvalues negative when $\alpha_{2}>3$ for $\gamma \neq0$ and hence behaves as an attracting node with $\omega_{eff}=-1$ and $\Omega_{tt}=1$. $F_{2}$ behaves as a late time attractor for $\alpha_{2}\in (2,3)$ which is stable and $F_{3}$ is also stable for $\alpha_{2}<3$ with $\omega_{eff}=-1$ and $\Omega_{tt}=1$. The presence of stable fixed points $ F_{2}$ and $F_{3}$ assures the presence of negative pressure and thereby contributes to the developed cosmological model with late time attractor solutions that represent the accelerated expansion phase of the Universe. The phase plot shown in Figs. (1) and (2) have supported these analytical results. With the notion of spectral radius we obtained a finer region of $\alpha_{2}$ where $F_{1}$ is stable, that is, $2<\alpha_{2}<3$. Fig. 3 shows that $F_{1}$ becomes saddle point for $\alpha_{2}>3$, $\gamma \neq 0$ where trajectories in some directions are attracted towards it and some trajectories along other directions are being repelled from it. The presence of $F_{1}$ in our system represents the stable dark energy model with $\omega=-1$ and $\Omega_{tt}=1$. We also obtain a non-hyperbolic fixed point $F_{2}$ which is stable for $\alpha_{2}\in [0,3)$. Fig. 4 and Fig. 5 also support this analytical findings. Stability for points at infinity has been analysed by using Poincar\'{e} sphere where we use stereographic projection to study the behavior of trajectories far from origin. The critical point occurs at the points $(\pm1,0,0)$ on the equator of the Poincar\'{e} sphere $S^{2}$. When $\gamma=0$, both eigen values $m_{1}$ and $m_{2}$ are negative for $\alpha_{2}<3$ and the critical point $(1,0,0)$ behaves as a stable attractor which represents the late time accelerated expansion phase of the Universe. For $\alpha_{2}>3$, both $m_{1}$ and $m_{2}$ are positive and the critical point $(1,0,0)$ behaves as an unstable repeller representing the inflationary epoch of the evolving Universe. Fig. 6 and Fig. 7 show the phase plot of the stable attractor and the unstable repeller respectively. For $\gamma=\frac{4}{3}$, $m_{1}>0$ and $m_{2}<0$ when $\alpha_{2}<3$ and the critical point $(1,0,0)$ behaves as a saddle point which is unstable representing the matter dominated phase of the evolving Universe. When $\alpha_{2}>3$, both $m_{1}$ and $m_{2}$ are positive and the critical point $(1,0,0)$ behaves as an unstable repeller. Fig. 8 and Fig. 9 also support the above analytical results for $\gamma=\frac{4}{3}$. For $\gamma=2$, the behavior is same as that of $\gamma=\frac{4}{3}$. Since the degree of the polynomial system $f(x,y)$ and $g(x,y)$ is odd, the behavior at the antipodal point $(-1,0,0)$ is exactly the same as the behavior at $(1,0,0)$. In case II of section 3, we present the case when $\rho_{\Lambda}=constant$ but $G$ no longer remains constant. By introducing new variables, we represent the model with a two dimensional dynamical system where we obtain two non-hyperbolic fixed points $P,Q$. We present the stability analysis of these fixed points by using spectral radius as well as perturbation function where we have found that both are stable for $\gamma\in [0,\frac{1}{3})$ with $\Omega_{tt}=-1$ and effective equation of state $\omega_{eff}=-1$. Also for $P$, both $\eta_{x}$ and $\eta_{y}$ converge to a constant value as $\Theta$ tends to infinity.  When $\gamma \neq 0$ $\eta_{y}\rightarrow -b$ as $\Theta \rightarrow \infty$ but if we directly put $\gamma=0$ in the expression of $\eta_{y}$, it becomes a constant function, that is, $\eta_{y}=c_{1}-b$. Fig. 10 shows the variation of perturbation along $y-$axis , $\eta_{y}$ with respect to $\Theta$ as $\gamma\rightarrow 0^{+}$ for the fixed point $P$. From Fig. 10 we see that as $\gamma\rightarrow 0$ from the right the curves gradually tends to $\eta_{y}=c_{1}-b$. For $Q$, $\eta_{x}$ evolves to a constant value and $\eta_{y}$ decays to zero as $\Theta$ gradually increases and tends to infinity as shown in Fig. 11 for $\gamma<\frac{1}{3}$. So both the fixed points are stable which gives the dark energy model which forms the strong base for the fact that the Universe is undergoing not just expansion but expansion with acceleration. When we take both $G$ and $\rho_{\Lambda}$ to be non-constants, then we see from case III of section 3 that we can extend the system to a three dimensional problem. We have analysed the system when $\alpha_{2}=3$ under three different values of $\gamma$, that is, $\gamma=0$(dark energy model),$\gamma=\frac{4}{3}$(radiation dominated model), $\gamma=2$(stiff fluid model) and study the system about its stability and corresponding cosmological implications. At $\gamma=0$ the fixed point $S$ is non-hyperbolic as some of the eigenvalues of the Jacobian matrix vanishes. Since $S$ is non-hyperbolic, we do the stability analysis by studying how the perturbation along each of the three axis vary with the increase in $\Theta$. As the set $\Phi=\phi$, $S$ is unstable. Fig. 12, Fig. 13, Fig. 14 shows the perturbation plots for $S$. We have also used Center manifold theory to analyze stability by using a suitable coordinate transformation where we obtain the standard form to apply Center manifold theory. As the dynamics of the center manifold is unstable we deduce that $S$ is unstable. From both approaches we find that $S$ is unstable. For $\gamma=\frac{4}{3}$ as well as $\gamma=2$, $S$ is non-hyperbolic and unstable. Fig. 15 and Fig. 16 show the perturbation plots of $S$ for $\gamma=\frac{4}{3}$. The perturbation function along each of the axis fail to decay or evolve to a constant value as $\Theta \rightarrow \infty$ which shows that $S$ is unstable. Fig. 17 shows that $\eta_{z}$ continues to increase exponentially as $\Theta$ increases which indicates that $S$ is unstable for $\gamma=2$ also. To analyse stability at infinity we use the concept of Poincar\'{e} sphere as any polynomial system in rectangular coordinates can be extended to the Poincar\'{e} sphere\cite{Roland}. Here since the system is a three dimensional system, the ideas of projective geometry has been carried over to higher dimension to analyse stability for flows in $\mathbb{R}^{3}$\cite{Poincare}. The critical points at infinity occur at the points $(\pm1,0,0,0)$ on the equator the Poincar\'{e} sphere $S^{3}$. Since the perturbation along each of the axis $\eta_{y}$, $\eta_{z}$ and $\eta_{w}$ decays to zero as cosmic time $\Theta$ tends to infinity as shown in Figs. 18, 19 and 20, we conclude that the fixed point $(\pm1,0,0,0)$ is a stable attractor. Throughout the entire work the developed cosmological models strongly support the fact that the Universe is in the phase of expansion with acceleration thereby depicting that our model has a deep connection with the accelerated expansion phenomena.\\

\section*{Declaration}
The authors declare that there is no conflict of interest regarding the publication of this paper.

\end{document}